\documentclass[11pt,twoside]{article}
\usepackage{epsfig}
\renewcommand\refname{REFERENCES}

\renewenvironment{thebibliography}[1]
 { \section*{\refname}
		\begin{list}{\arabic{enumi}.}
    {\usecounter{enumi} \setlength{\parsep}{0pt}
     \setlength{\itemsep}{3pt} \settowidth{\labelwidth}{#1.}
     \sloppy
    }}{\end{list}}
\begin{document}
%%%%%%%%%%%%%%%%%%%%%%%%%%%%%%%%%%%%%%%%%%%%%%%%%%%%%%%%%%%%%%%%%%%%%%%%%%%%%%%%%%%

\title{On the measurement problem for a two-level quantum system }
\author{Alexey A. Kryukov
\footnote{Department of Mathematics, University of Wisconsin Colleges 
\newline
E-mail: alexey.kryukov@uwc.edu, aakrioukov@facstaff.wisc.edu}
}
\maketitle

\pagestyle{myheadings} 
\thispagestyle{plain}         
\markboth{Alexey A. Kryukov}{On the measurement problem} 
\setcounter{page}{1}

%\vskip60pt
\begin{abstract}
A geometric approach to quantum mechanics with unitary evolution and non-unitary collapse processes is developed. In this approach the Schr{\"o}dinger evolution of a quantum system is a geodesic motion on the space of states of the system furnished with an appropriate Riemannian metric. The measuring device is modeled by a perturbation of the metric. The process of measurement is identified with a geodesic motion of state of the system in the perturbed metric. Under the assumption of random fluctuations of the perturbed metric, the Born rule for probabilities of collapse is derived. The approach is applied to a two-level quantum system to obtain a simple geometric interpretation of quantum commutators, the uncertainty principle and Planck's constant. In light of this, a lucid analysis of the double-slit experiment with collapse and an experiment on a pair of entangled particles is presented.   
\end{abstract}


\bigskip

%%%%%%%%%%%%%%%%%%%%%%%%%%%%%%%%%%%%%%%%%%%%%%%%%%%%%%%%%%%%%%%%%

\section{GEOMETRY AND QUANTUM MECHANICS}
\setcounter{equation}{0}

Geometric ideas have played a well recognized role in modern physics, especially in general relativity (GR) and gauge theories (GT). They also found a well established position in quantum mechanics (QM) in considerations related to Berry's phase Ref. \cite{Berry}. However, whereas in GR and GT geometry (i.e., the metric or connection) defines the dynamics of the theory, the geometric methods pertaining to Berry's phase do not enjoy such a sweeping significance. The reason for this difference is quite obvious. Indeed, the geometry underlying GR and GT is directly related to the physical fields (gravitational or gauge) in the theory. At the same time, the Fubini-Study metric in the geometric interpretation of Berry's phase (Refs. \cite{AA},\cite{Simon} amongst many others) depends only on the geometry of the Hilbert space of states of quantum system. The latter geometry (i.e. Hilbert metric) is insensitive to changes in the Hamiltonian of the system and, consequently, is not dynamical.

At the same time, it turns out to be easy to make the metric on Hilbert space of states of a closed quantum system dynamical Refs. \cite{Kryukov},\cite{Kryukov1}. For this, notice first of all that the Schr{\"o}dinger equation
\begin{equation}
\label{Schroed}
\frac{d \varphi_{t}}{dt}=-\frac{i}{{\hbar}}{\widehat h}\varphi_{t}
\end{equation}
is the equation for integral curves of the vector field $h_{\varphi}:H \longrightarrow TH$, $h_{\varphi}=-\frac{i}{{\hbar}}{\widehat h}\varphi$ associated with the Hamiltonian ${\widehat h}$ of the system. Here $H$ is the Hilbert space of states of the system and $TH$ is the tangent bundle over $H$. Assume that the Hilbert space $H$ is a space of functions that are square-integrable with respect to an appropriate measure. Because the evolution governed by Eq. (\ref{Schroed}) is unitary, the integral curve through initial point $\varphi_{0}$ on the unit sphere $S^{H}$ in $H$ will stay on the sphere. Since this holds true for any initial point (modulo the domain issues), one concludes that the restriction of the vector field $h_{\varphi}$ to the sphere $S^{H}$ is a vector field on the sphere.

In the ordinary QM spaces $T_{\varphi}H$ tangent to $H$ at $\varphi \in H$ are identified with the space $H$ itself. Similarly, spaces $T_{\varphi}S^{H}$ tangent to the sphere $S^{H}$ at $\varphi \in S^{H}$ are identified with (real) affine subspaces of $H$. In particular, the metric on $S^{H}$, whenever used, is assumed to be induced by the embedding of $S^{H}$ into $H$. 

However, the sphere $S^{H}$ is a manifold and thus, can be defined independently of the ambient space $H$. As such, $S^{H}$ is a {\em Banach manifold} which means that it can be obtained by ``gluing together'' open sets in a Banach space.  
The Hilbert metric $G_{\varphi}:T_{\varphi}H \times T_{\varphi}H \longrightarrow \bf {C}$ on tangent spaces $T_{\varphi}H$ can be also defined independently of the metric on $H$ as an Hermitian tensor field on $H$. Such a tensor field gives rise to a Riemannian metric $G_{R\varphi}:T_{\varphi}S^{H} \times T_{\varphi}S^{H} \longrightarrow \bf {R}$ on $S^{H}$, defined at each $\varphi \in S^{H}$ by 
\begin{equation}
G_{R\varphi}(X,Y)=2Re G_{\varphi}(\xi, \eta).
\end{equation}
Here $X=(\xi, {\overline \xi}), Y=(\eta, {\overline \eta}) \in T_{\varphi}S^{H}$ are considered as vectors in the realization of the tangent space $T_{\varphi}H$.
The manifold $S^{H}$, furnished with the $(2,0)$-tensor field $G_{R\varphi}$, is then a {\em Riemannian manifold}.
In the following, the manifold $S^{H}$ with the metric $G_{R\varphi}$ will be denoted by $S^{G}$.

The final step in making the metric $G_{\varphi}$ on the sphere of states $S^{G}$ dynamical is to ensure that 
the integral curves of $h_{\varphi}$ (i.e. the solutions to Schr{\"o}dinger equation Eq. (\ref{Schroed})) are geodesics on $S^{G}$. For this it turns out to be sufficient to define $G_{\varphi}:T_{\varphi}H \times T_{\varphi}H \longrightarrow \bf {C}$ by
\begin{equation}
\label{metric}
G_{\varphi}(\xi, \eta)=\hbar^{2}\left(({\widehat h}{\widehat h}^{\ast})^{-1}\xi, \eta \right)_{H}.
\end{equation}
Here ${\widehat h}^{\ast}$ is the adjoint of ${\widehat h}$ (normally equal to ${\widehat h}$) and the Hamiltonian ${\widehat h}$ is assumed to be invertible. Incidentally, even if the Hamiltonian ${\widehat h}$ is not bounded on $H$, it becomes bounded as an operator mapping points $\varphi \in H$ into tangent spaces $T_{\varphi}H$ with the metric Eq. (\ref{metric})  (see Ref. \cite{Kryukov1}).

Further general results concerning QM on Hilbert manifolds can be found in Ref. \cite{Kryukov}. 
%These results strongly support the point of view  that the space of states is a new arena for physical processes. The goal of this Letter is to demonstrate the advantages of such a point of view by applying it to a simple two-level system.
These results demonstrate that QM can be formulated in terms of geometry of the space of states. 
The goal of the Letter is to provide such a geometric formulation in case of a simple two-level system and to establish its advantages. Namely, the point of view that the space of states represents a new arena for physical processes and the evolution of state is a motion along geodesic is shown to be effective in addressing the major conceptual difficulties of quantum mechanics. Although the discussion deals primarily with a simple model, the most important results can be shown to be quite general. Some of these generalizations are described in the Letter while others are left for the upcoming publications.

\section{ELECTRON IN A HOMOGENEOUS MAGNETIC FIELD}
\setcounter{equation}{0}
 
Consider a free non-relativistic electron propagating in the direction of the $X$-axis in a homogeneous magnetic field ${\bf B}$. The evolution equation (the Pauli equation) for the electron is
\begin{equation}
\label{Pauli}
i\hbar \frac{d \Psi}{dt}=-\frac{\hbar^{2}}{2m}\frac{d^{2}}{dx^{2}} \Psi-\mu {\widehat {\bf \sigma}}\cdot {\bf B}\Psi,
\end{equation}  
where $\Psi=\Psi(s,x,t)$, $s=1,2$ is a two-component state function of the electron, $\mu$ is the electron's magnetic moment and ${\bf {\widehat \sigma}}=({\widehat \sigma}_x, {\widehat \sigma}_{y}, {\widehat \sigma}_{z})$ is the vector made of Pauli matrices. The substitution
$\Psi(s,x,t)=\psi_{t}(x)\varphi_{t}(s)$
separates variables and produces two independent evolution equations. The first describes the evolution governed by the free Hamiltonian
\begin{equation}
\label{freeeH}
i\hbar \frac{d \psi_{t}}{dt}=-\frac{\hbar^{2}}{2m}\frac{d^{2}}{dx^{2}} \psi_{t}.
\end{equation}
The second equation describes the evolution in the space $C^{2}$ of spinors $\varphi$:
\begin{equation}
\label{spin_ev}
i\hbar \frac{d \varphi_{t}}{dt}=-\mu {\widehat {\bf \sigma}} \cdot {\bf B}\varphi_{t}.
\end{equation}
It follows that in the case of the product states $\Psi(s,x,t)=\psi_{t}(x)\varphi_{t}(s)$, one can analyze the evolution of spin state $\varphi_{t}$ in the space of states $C^{2}$ without needing to involve the infinite-dimensional Hilbert space of states $\Psi$.

\subsection{Quantum Mechanics on the Space of States $S^{3}$}

Let us proceed to reformulation of quantum mechanics of the system in geometrical terms. 
In this, the fact that the sphere $S^{3}$ of unit normalized spin states can be furnished with the group structure of the group $SU(2)$ will be helpful. The group structure will allow us to exploit simple results from differential geometry of Lie groups which will make the resulting picture more transparent and complete. 

First of all, the Hamiltonian ${\widehat h}=-\mu {\widehat {\bf \sigma}} \cdot {\bf B}$ defines the vector field $h_{\varphi}=\frac{i}{\hbar}\mu {\widehat {\bf \sigma}} \cdot {\bf B}\varphi$ on the sphere $S^{3}$ in the space $C^{2}$ with the metric $(\xi, \eta)_{C^{2}}=\sum_{k} \xi_{k} {\overline \eta}_{k}$. The integral curve of $h_{\varphi}$ (i.e. the solution of Eq. (\ref{spin_ev})) through $\varphi_{0} \in S^{3}$ is given by 
\begin{equation}
\label{int_curve}
\varphi_{t}=e^{\frac{i}{\hbar}\mu {\widehat {\bf \sigma}} \cdot {\bf B}t}\varphi_{0}.
\end{equation}
Since $\varphi_{t}$ is a path in $C^{2}$, it is natural to call the vector $\frac{d \varphi_{t}}{dt}$ the {\it velocity of evolution} of the system. The {\it speed of evolution} in the $C^{2}$ is the norm of $\frac{d \varphi_{t}}{dt}$ in $C^{2}$ metric. 
Using 
\begin{equation}
\label{sigma}
({\bf {\widehat \sigma}} \cdot {\bf A})({\bf {\widehat \sigma}} \cdot {\bf B})={\bf A}\cdot {\bf B}+i{\bf {\widehat \sigma}} \cdot {\bf A} \times {\bf B},
\end{equation} 
one has
\begin{equation}
\label{sigma1}
({\bf {\widehat \sigma}} \cdot {\bf B})^{2}={\bf B}^{2}.
\end{equation}
Therefore, by Hermicity of the matrix ${\widehat {\bf \sigma}} \cdot {\bf B}$, one obtains 
\begin{equation}
\label{speed}
\left\| \frac{d \varphi_{t}}{dt} \right \|_{C^{2}}=\left(\frac{i}{\hbar}\mu {\widehat {\bf \sigma}} \cdot {\bf B}\varphi_{t}, \frac{i}{\hbar}\mu {\widehat {\bf \sigma}} \cdot {\bf B}\varphi_{t}\right)^{\frac{1}{2}}_{c^{2}}=\frac{\mu B}{\hbar},
\end{equation}
where $B$ is the norm of ${\bf B}$. In particular, the speed of evolution of the system depends only on the magnitude of the field.

To make the evolution of the system a motion along a geodesic, the metric $G_{\varphi}$ on $S^{3}$ will be defined by Eq. (\ref{metric}). Since ${\widehat h}$ is self-adjoint, one obtains ${\widehat h}{\widehat h}^{\ast}={\widehat h}^{2}=\mu^{2}({\bf {\widehat \sigma}} \cdot {\bf B})^{2}=\mu^{2}B^{2}I$, where $I$ is the identity operator on $C^{2}$ and  Eq. (\ref{sigma1}) has been used at the last step. Therefore, up to the constant factor $(\hbar/\mu B)^{2}$, the metric $G_{\varphi}$ coincides with the one induced by the embedding of $S^{3}$ into $C^{2}$. That means that the carriers of the geodesics on $S^{3}$ are the intersections of $S^{3}$ with the planes through the origin. The fact that the found Riemannian metric is so simple is due to an especially simple form of the Hamiltonian in the model.

If $S^{3}$ is identified with the group manifold $SU(2)$, the obtained metric is the Killing metric on $SU(2)$. To see this, let us identify in the standard way the space $C^{2}$ of complex vectors $\varphi=\left[ \begin{array}{c}
z_{1} \\ 
z_{2}
\end{array}
\right]$ 
with the space $Mat$ of $2 \times 2$ matrices 
\begin{equation}
\label{Mat}
{\widehat \varphi}=\left[ 
\begin{array}{cc}
z_{1} & z_{2} \\ 
-{\overline z}_{2} & {\overline z}_{1}
\end{array}
\right].
\end{equation}
The map $\omega: \varphi \longrightarrow {\widehat \varphi}$ is an isomorphism of (real) vector spaces $C^{2}$ and $Mat$.
The sphere $S^{3}$ of unit states in $C^{2}$ is identified via $\omega$ with the subset of matrices with unit determinant. The latter subset is the group $SU(2)$ under matrix multiplication. 

The Killing metric on the Lie algebra $su(2)$ can be defined by
\begin{equation}
\label{su2}
({\widehat X}, {\widehat Y})_{K}= c Tr ({\widehat X}{\widehat Y}^{+}),
\end{equation}
where $c\neq 0$ is an arbitrary constant and $Tr$ stands for the trace. 
The Killing metric on the group $SU(2)$ is defined for all ${\widehat \varphi} \in SU(2)$ and all left-invariant vector fields $L_{\widehat X}({\widehat \varphi})={\widehat \varphi}{\widehat X}, L_{\widehat Y}({\widehat \varphi})={\widehat \varphi}{\widehat Y}$, ${\widehat X}, {\widehat Y} \in su(2)$ by
\begin{equation}
\label{Killing}
(L_{\widehat X}({\widehat \varphi}), L_{\widehat Y}({\widehat \varphi}))_{K}=(L_{\widehat X}(e), L_{\widehat Y}(e))_{K}=({\widehat X}, {\widehat Y})_{K},
\end{equation}
where $e$ is the identity element in $SU(2)$. 

By direct substitution one verifies that $(L_{\widehat X}(e), L_{\widehat Y}(e))_{K}=2 c Re (X,Y)_{C^{2}}$ whenever ${\widehat X}=d\omega (X)$ and ${\widehat Y}=d\omega (Y)$ with $d\omega$ being the differential of the map $\omega$. In other words, the Killing metric is proportional to the metric induced by the embedding of $S^{3}$ into the Euclidean space $C^{2}=R^{4}$. This verifies that the metric $G_{\varphi}=(\hbar /\mu B)^{2} I$ obtained earlier, is the Killing metric.

For any two left invariant vector fields $L_{{\widehat X}}, L_{{\widehat Y}}$ on $SU(2)$ the connection $\nabla$ on $SU(2)$ can be defined by
\begin{equation}
\label{nabla}
\nabla_{L_{{\widehat X}}}L_{{\widehat Y}}=\frac{1}{2}L_{[{\widehat X},{\widehat Y}]}.
\end{equation}
Notice that the left invariant vector fields form a basis on the tangent space $T_{{\widehat \varphi}}SU(2)$ for all ${\widehat {\varphi}} \in SU(2)$. In particular, Eq. (\ref{nabla}) is sufficient to define a connection on $SU(2)$. This connection is symmetric, as the torsion tensor vanishes: 
\begin{equation}
T(L_{{\widehat X}}, L_{{\widehat Y}})=\nabla_{L_{{\widehat X}}}L_{{\widehat Y}}-\nabla_{L_{{\widehat Y}}}L_{{\widehat X}}-[L_{{\widehat X}}, L_{{\widehat Y}}]=\frac{1}{2}L_{[{\widehat X},{\widehat Y}]}-\frac{1}{2}L_{[{\widehat Y},{\widehat X}]}-L_{[{\widehat X},{\widehat Y}]}=0.
\end{equation}
The connection Eq. (\ref{nabla}) is also compatible with the Killing metric, that is, for any vector fields $\xi, \eta, \zeta$ on $SU(2)$ the following is true:
\begin{equation}
\label{compatible}
\nabla_{\xi}(\eta, \zeta)_{K}=(\nabla_{\xi}\eta, \zeta)_{K}+(\eta, \nabla_{\xi}\zeta)_{K}.
\end{equation}
Indeed, assuming that $\xi=L_{{\widehat X}}, \eta=L_{{\widehat Y}}, \zeta=L_{{\widehat Z}}$ are left invariant, one has 
\begin{equation}
(L_{{\widehat Y}}({\widehat \varphi}), L_{{\widehat Z}}({\widehat \varphi})_{K}=(L_{\widehat Y}(e), L_{\widehat Z}(e))_{K}=const 
\end{equation}
and therefore the left hand side of Eq. (\ref{compatible}) vanishes. For the right hand side, by definition Eq. (\ref{nabla}) one obtains:
\begin{equation}
(\nabla_{L_{{\widehat X}}}L_{{\widehat Y}}, L_{{\widehat Z}})_{K}+(L_{{\widehat Y}}, \nabla_{L_{{\widehat X}}}L_{{\widehat Z}})_{K}=\frac{1}{2}([{\widehat X}, {\widehat Y}], {\widehat Z})_{K}+\frac{1}{2}({\widehat Y}, [{\widehat X}, {\widehat Z}])_{K}.
\end{equation}
From the anti-Hermicity of elements of $su(2)$ one also has:
\begin{equation}
([{\widehat X}, {\widehat Y}], {\widehat Z})_{K}=-c Tr({\widehat X}{\widehat Y}{\widehat Z})+c Tr({\widehat Y}{\widehat X}{\widehat Z})
\end{equation}
and
\begin{equation}
([{\widehat Y}, {\widehat X}], {\widehat Z})_{K}=-c Tr({\widehat Y}{\widehat X}{\widehat Z})+c Tr({\widehat X}{\widehat Y}{\widehat Z}).
\end{equation}
As a result, the sum on the right hand side of Eq. (\ref{compatible}) is also zero which verifies that the connection Eq. (\ref{nabla}) is compatible with the metric. In other words, the connection $\nabla$ is the Levi-Civita connection of the Killing metric.

For any magnetic field ${\bf B}$ in the model the one parameter subgroup ${\widehat \varphi}_{t}=e^{-\frac{i}{\hbar}\mu {\widehat {\bf \sigma}} \cdot {\bf B}t}$ is a geodesic through the identity $e \in SU(2)$. Indeed, since $\frac{d{\widehat \varphi}_{t}}{dt}=-{\widehat \varphi}_{t}\frac{i}{\hbar}\mu {\widehat {\bf \sigma}} \cdot {\bf B}$, the path ${\widehat \varphi}_{t}$ is the integral curve of the left invariant vector field $L_{\widehat h}{{\widehat \varphi}}=-{\widehat \varphi}\frac{i}{\hbar}\mu {\widehat {\bf \sigma}} \cdot {\bf B}$. Using the definition Eq. (\ref{nabla}) one then has:
\begin{equation}
\nabla_{\frac{d{\widehat \varphi}_{t}}{dt}}\frac{d{\widehat \varphi}_{t}}{dt}=\nabla_{L_{\widehat h}}L_{{\widehat h}}=\frac{1}{2}L_{[{\widehat h}, {\widehat h}]}=0.
\end{equation}
Geodesics through an arbitrary point ${\widehat \varphi}_{0} \in SU(2)$ can be then written in the form ${\widehat \varphi}_{t}={\widehat \varphi}_{0}e^{-\frac{i}{\hbar}\mu {\widehat {\bf \sigma}} \cdot {\bf B}t} $. Considered as paths with values in $C^{2}$, these geodesics take the form $\varphi_{t}=e^{\frac{i}{\hbar}\mu {\widehat {\bf \sigma}} \cdot {\bf B}t}\varphi_{0}$.

The curvature tensor of $\nabla$ can be obtained directly from the definition
\begin{equation}
R(L_{{\widehat X}},L_{{\widehat Y}})L_{{\widehat Z}}=\nabla_{L_{{\widehat Y}}}\nabla_{L_{{\widehat X}}}L_{{\widehat Z}}-\nabla_{L_{{\widehat X}}}\nabla_{L_{{\widehat Y}}}L_{{\widehat Z}}+\nabla_{[L_{{\widehat X}},L_{{\widehat Y}}]}L_{{\widehat Z}}.
\end{equation}
In particular, 
\begin{equation}
\label{curv1SU2}
R(L_{{\widehat X}},L_{{\widehat Y}})L_{{\widehat Z}}=\frac{1}{4}L_{[[{\widehat X},{\widehat Y}],{\widehat Z}]}.
\end{equation}
and
\begin{equation}
\label{curv2SU2}
\left(R(L_{{\widehat X}},L_{{\widehat Y}})L_{{\widehat Z}},L_{{\widehat W}}\right)_{K}=\frac{1}{4}\left([{\widehat X},{\widehat Y}],[{\widehat Z},{\widehat W}]\right)_{K}.
\end{equation}
The sectional curvature in the plane through $L_{\widehat X}, L_{\widehat Y}$ is defined by
\begin{equation}
\label{sectional}
\frac{\left(R(L_{\widehat X},L_{\widehat Y})L_{\widehat X},L_{\widehat Y}\right)_{K}}{\left(L_{\widehat X},L_{\widehat X}\right)_{K}\left(L_{\widehat Y},L_{\widehat Y}\right)_{K}-\left(L_{\widehat X},L_{\widehat Y}\right)_{K}}.
\end{equation}
With the help of Eqs. (\ref{curv2SU2}) and (\ref{Killing}) this becomes
\begin{equation}
\label{sectional1}
\frac{1}{4}\frac{\left([{\widehat X},{\widehat Y}],[{\widehat X},{\widehat Y}]\right)_{K}}{\left({\widehat X},{\widehat X}\right)_{K}\left({\widehat Y},{\widehat Y}\right)_{K}-\left({\widehat X},{\widehat Y}\right)_{K}}.
\end{equation}

Suppose for example that ${\widehat X}$, ${\widehat Y}$, ${\widehat Z}$ and ${\widehat W}$ correspond to the spin observables.
Recall that in the Planck system of units the operator of spin ${\bf {\widehat s}}$ has eigenvalues $\pm 1/2$ and can be expressed in terms of the Pauli matrices ${\widehat \sigma}_{1}, {\widehat \sigma}_{2}, {\widehat \sigma}_{3}$ as
\begin{equation}
{\bf {\widehat s}}=\frac{1}{2}{\bf {\widehat \sigma}},
\end{equation}
where ${\bf {\widehat \sigma}}=({\widehat \sigma}_{1}, {\widehat \sigma}_{2}, {\widehat \sigma}_{3})$. 
The corresponding anti-Hermitian generators ${\widehat e}_{k}=\frac{i}{2} {\widehat \sigma}_{k}$, $k=1,2,3$, form a basis of the Lie algebra $su(2)$ and satisfy the commutator relations
\begin{equation}
\left[{\widehat e}_{k},{\widehat e}_{l}\right]= \epsilon _{klm}{\widehat e}_{m},
\end{equation}
where $\epsilon_{klm}$ denotes the completely antisymmetric tensor of rank three.  
In the basis $\{{\widehat e}_{k}\}$ the curvature tensor Eqs. (\ref{curv1SU2}), (\ref{curv2SU2}) takes the form
\begin{equation}
R^{i}_{k,lm}=\frac{1}{4}(\delta^{i}_{l}\delta_{km}-\delta^{i}_{m}\delta_{kl}),
\end{equation}
and
\begin{equation}
R_{ik,lm}=\frac{c}{8}(\delta_{il}\delta_{km}-\delta_{im}\delta_{kl})
\end{equation}
where $\delta_{ik}$ is the Kronecker delta.
The symmetry property
\begin{equation}
R^{i}_{k,lm}+R^{i}_{l,mk}+R^{i}_{m,kl}=0
\end{equation}
of the curvature tensor coincides in the model with the Jacobi identity
\begin{equation}
[[{\widehat X}, {\widehat Y}], {\widehat Z}]+[[{\widehat Y}, {\widehat Z}], {\widehat X}]+[[{\widehat Z}, {\widehat X}], {\widehat Y}]=0
\end{equation}
for the Lie algebra elements ${\widehat X}, {\widehat Y}, {\widehat Z}$.

From the isomorphism given by Eq. (\ref{Mat}) it follows that any vector ${\bf x}=(x^{1}, x^{2}, x^{3})$ in the Euclidean space $R^{3}$ can be identified with the element $\sum_{k} x^{k}i{\widehat \sigma}_{k}=\sum_{k} 2x^{k}{\widehat e}_{k}$ of the Lie algebra $su(2)$. Under such an identification the Euclidean norm $\left\|{\bf x}\right\|_{R^{3}}$ of ${\bf x}$ is equal to $\det {\bf x}$ and rotations in $R^{3}$ are represented by transformations ${\bf x} \longrightarrow {\widehat U}{\bf x}{\widehat U}^{+}$ with ${\widehat U} \in SU(2)$. 
One can make this identification into an isometry by assuming the equality of Euclidean and Killing norms
\begin{equation}
\label{fixing}
\left \|\sum_{k} 2x^{k}{\widehat e}_{k}\right \|_{K}=\left \|{\bf x}\right \|_{R^{3}}.
\end{equation}
This will fix the constant factor in front of the Killing metric. 

Note that the Euclidean space $R^{3}$ can be identified with the space of all possible classical angular momenta of a particle. The electron's possible angular momenta form a sphere $S^{2}$ in $R^{3}$. By equating the norms according to Eq. (\ref{fixing}), the spaces tangent to $S^{2}$ are identified with affine subspaces of spaces tangent to the sphere of states $S^{3}$ with the induced metric. In particular, the sphere $S^{2}$ can be identified with a submanifold of the space of states $S^{3}$ with the induced metric. Let us remark that this identification is analogous to the identification of the classical space with the submanifold of point supported states in an infinite-dimensional Hilbert space of states, considered in Refs. \cite{Kryukov}, \cite{Kryukov3}. 

In the Killing metric Eq. (\ref{Killing}) on $S^{3}$ one has $\left(\sum_{k} 2x^{k}{\widehat e}_{k}, \sum_{m} 2x^{m}{\widehat e}_{m}\right)_{ K}=2c \sum_{k,m} x^{k}x^{m}\delta_{km}$. To satisfy Eq. (\ref{fixing}) the constant $c$ must be $1/2$, that is, the needed metric in Planck units has the form
\begin{equation}
\label{norm_metric}
({\widehat X}, {\widehat Y})_{K}= \frac{1}{2}Tr ({\widehat X}{\widehat Y}^{+}).
\end{equation}
Using the formula Eq. (\ref{sectional1}), one obtains the following expression for the sectional curvature 
$R(p)$ in the plane $p$ through orthogonal vectors $L_{{\widehat e}_{1}},L_{{\widehat e}_{2}}$: 
\begin{equation}
\label{section}
\frac{1}{4} \frac{ \left([{\widehat e_{1}},{\widehat e_{2}}],[{\widehat e_{1}},{\widehat e_{2}}]\right)_{K}}{\left({\widehat e}_{1},{\widehat e}_{1}\right)_{K}\left({\widehat e}_{2},{\widehat e}_{2}\right)_{K}}=4\left ({\widehat e}_{3},{\widehat e}_{3}\right)_{K}=1.
\end{equation}
So the sectional curvature of $S^{3}$ in Planck units is equal to $1$. 
Note that in an arbitrary system of units the Killing metric would be multiplied by $\hbar^{2}$ and the sectional curvature would be equal to $1/ \hbar^{2}$. The dimension of sectional curvature is consistent with the fact that the tangent space $su(2)$ is spanned by the spin operators having the dimension of angular momentum.

In this approach the Planck's constant and the commutators of spin observables acquire a transparent geometric interpretation. According to Eq. (\ref{sectional1}) the commutator of two observables is directly related to the sectional curvature of the sphere $S^{3}$. 
Indeed, assume for simplicity that the vector fields $L_{{\widehat X}}, L_{{\widehat Y}}$ are orthogonal and unit normalized in the Killing metric. Then from Eqs. (\ref{sectional}), (\ref{sectional1}) one has the following expression for the norm of the commutator of  ${\widehat X}$ and  ${\widehat Y}$:
\begin{equation}
\label{normm}
\left\|[{\widehat X}, {\widehat Y}]\right\|^{2}_{K}=4 R(p).
\end{equation}
Here $R(p)$ is the sectional curvature of $S^{3}$ in the plane $p=\mathcal{L}(L_{\widehat X},L_{\widehat Y})$ which for the considered Riemannian metric was found to be a constant equal to $1/\hbar^{2}$. 

Note that for $L_{{\widehat X}}, L_{{\widehat Y}}$ which are orthogonal but not unit, the equation Eq. (\ref{sectional1}) takes the form $\left\|[{\widehat X}, {\widehat Y}]\right\|^{2}_{K}=4 R(p)\left\|{\widehat X}\right\|^{2}_{K} \left\|{\widehat Y}\right\|^{2}_{K}$. In particular, if the norms of ${\widehat X}$ and ${\widehat Y}$ are of order $\hbar$ (e.g., ${\widehat X}, {\widehat Y}$ are the spin observables), then the norm of the commutator $[{\widehat X}, {\widehat Y}]$ is of order $\hbar$ as well. Note that despite the fact that the commutator $[{\widehat X}, {\widehat Y}]$ is small in these units, it is of the order of the radius of the sphere of states, making quantum effects on the sphere quite transparent. 

The results obtained so far in this section were model specific. It is then important to know whether they can be generalized to the case of higher dimensional spaces of states and of arbitrary observables. Also, what if the Hamiltonian of the system is time-dependent? Here is a sketch of what can be done in these cases.

For any $n$ the sphere of states $S^{2n-1}$ in the space $C^{n}$ is a homogeneous space $U(n)/U(n-1)$, where $U(n)$ denotes the unitary group on $C^{n}$. The Killing metric on $U(n)$ can be used to induce a Riemannian metric on $S^{2n-1}$ via the embedding. Namely, the $2n-1$ linearly independent generators in the Lie algebra $u(n)$, which belong to the orthogonal complement of a (fixed) subalgebra $u(n-1)$, form a subspace $V \subset u(n)$. The one-parameter subgroups $e^{{\widehat X}\tau}$ with ${\widehat X}$ in $V$ sweep a sphere $S^{2n-1}$ and yield geodesics in the induced metric. The curvature of $S^{2n-1}$ can be then computed via equations Eqs. (\ref{curv1SU2}), (\ref{curv2SU2}) with generators in $V$. Furthermore, the commutators of generators in $V$ are related to the sectional curvature of $S^{2n-1}$ by the same formula Eq. (\ref{normm}).

Note that there exist anti-Hermitian observables that are not in $V$. For example, if $n=2$ so that $S^{3}=U(2)/U(1)=SU(2)$, then $V$ is the Lie subalgebra $su(2)\subset u(2)$ and so $V$ consists of the traceless elements of $u(2)$. If ${\widehat X} \in u(2)$ is not traceless, the one-parameter subgroup $e^{{\widehat X}\tau}$ is still a geodesic in $U(2)$. However, this geodesic does not ``stay'' on the subgroup $SU(2)=S^{3}$. Of course, one could still consider the curves on $S^{3}$ given by $\varphi_{t}=e^{{\widehat X}\tau}\varphi_{0}$, for some point $\varphi_{0}$ in $S^{3} \subset C^{2}$. For any given ${\widehat X} \in u(2)$ and all initial points $\varphi_{0}$ these curves are still geodesics in the appropriate Riemannian metric (see section 1 and Ref. \cite{Kryukov}). However, the algebraic features of the model change and the formulas connecting the commutators with the curvature are different.

For the time-dependent Hamiltonians ${\widehat h}$ in the Hilbert space $C^{n}$ the approach can be generalized as follows. The sphere of states $S^{2n-1}$ is replaced with the manifold $M=S^{2n-1} \times R$, where $R$ is the time line. Then, there exists a Riemannian metric on $M$ in which the paths $(\varphi_{t},t)$ with $\varphi_{t}=e^{-i{\widehat h}t}\varphi_{0}$ are geodesics for all $\varphi_{0} \in S^{2n-1}$. 

%Moreover, for most Hamiltonians the Riemannian metric on $M$ can be chosen in such a way that the sections $t=t_{0}$ have a metric originating in the Killing metric on the group $U(n)$.

Finally, as already mentioned, in the infinite-dimensional case there still exists a Riemannian metric for which all solutions to the Schr{\"o}dinger equation with an invertible (time-independent) Hamiltonian are geodesics. However, the algebraic properties of the model require further investigation in this case.

\subsection{Quantum Mechanics on the Projective Space of States $CP^{1}$}

In physical experiments one can only determine the state of a system up to a complex non-zero factor. That means that the space of physical states is the complex projective space $CP^{H}$ of complex lines in the space of states. In the considered example it is the one dimensional complex projective space $CP^{1}$. By definition, $CP^{1}$ is the quotient  $C^{2}_{\ast}/C_{\ast}$, were $\ast$ means ``take away zero''. In other words, $CP^{1}$ is the base manifold of the fibre bundle $\pi: C_{\ast}^{2} \longrightarrow CP^{1}$ with the natural projection along the fibres $C_{\ast}$. By considering unit normalized states only, one obtains $CP^{1}$ as a quotient $S^{3}/S^{1}$. It is then the base of the fibre bundle $\pi: S^{3} \longrightarrow CP^{1}$, which is a sub-bundle of the previous fibre bundle.

If $\varphi_{t}$ is a path of the electron's state on $S^{3}$ and $\pi: S^{3} \longrightarrow CP^{1}$ is the bundle projector, then $\pi(\varphi_{t})$ is a path on the base $CP^{1}$. Since this latter path represents what can be measured in experiments, it is important to obtain an explicit formula for $\pi$. For this consider a point 
$\varphi=\left[ \begin{array}{c}
\varphi_{1} \\ 
\varphi_{2}
\end{array}
\right]$
on $S^{3}$ and let 
$\{\varphi\}$ be the complex line formed by vectors $\lambda \varphi, \lambda \in C$. Provided $\varphi_{1} \neq 0$, there is a unique point of intersection of the line with the affine plane in $C^{2}$ formed by vectors
$\left[ \begin{array}{c}
1 \\ 
\xi
\end{array}
\right], \xi \in C$. Namely, by setting 
\begin{equation}
\lambda \left[ \begin{array}{c}
\varphi_{1} \\ 
\varphi_{2}
\end{array}
\right]=\left[ \begin{array}{c}
1 \\ 
\xi
\end{array}
\right],
\end{equation}
one obtains 
\begin{equation}
\label{xi}
\xi=\frac{\varphi_{2}}{\varphi_{1}}. 
\end{equation}
The map $\rho={\varphi} \longrightarrow \xi$ provides a coordinate chart on $CP^{1}$ which identifies $CP^{1}$ without a point (complex line through
$\left[ \begin{array}{c}
0 \\ 
1
\end{array}
\right]$)
with the set $C$ of complex numbers. The affine plane of vectors
$\left[ \begin{array}{c}
1 \\ 
\xi
\end{array}
\right]$ form a subspace in the Lie algebra $su(2)$. The algebra $su(2)$ itself has been identified earlier with the Euclidean space $R^{3}$ of vectors ${\bf x}= \sum_{k} x^{k} i{\widehat \sigma}_{k}$. The stereographic projection then identifies the unit sphere $S^{2}$ at the origin of $R^{3}$ with the above plane $C$ plus a point, i.e., with $CP^{1}$ itself.

The relationship of the coordinate $\xi$ in the plane $C$ with coordinates $(x^{1}=x,x^{2}=y,x^{3}=z)$ of the corresponding point on the sphere $S^{2}$ is given by
\begin{equation}
\xi=\frac{x+iy}{1-z}.
\end{equation}
Solving this for $x,y$ and $z$ and using Eq. (\ref{xi}), one obtains:
\begin{eqnarray}
\label{xyz}
x&=&\varphi_{1}{\overline \varphi_{2}}+{\overline \varphi_{1}}\varphi_{2},\\
\label{xyz1}
y&=&i(\varphi_{1}{\overline \varphi_{2}}-{\overline \varphi_{1}}\varphi_{2}), \\
\label{xyz2}
z&=&\varphi_{2}{\overline \varphi_{2}}-\varphi_{1}{\overline \varphi_{1}}.
\end{eqnarray}
The resulting map $\pi: S^{3} \longrightarrow S^{2}$ given by $(\varphi_{1},\varphi_{2}) \longrightarrow (x,y,z)$ is the needed projection on the space of physical states. 

The equation for the integral curve Eq. (\ref{int_curve}) can be simplified by choosing the coordinate axes properly. In particular, one can always assume that the $Z$-axis is parallel to the magnetic field ${\bf B}$. In this case ${\bf {\widehat \sigma}}\cdot {\bf B}={\widehat \sigma}_{3}B$ and Eq. (\ref{int_curve}) simplifies to 
\begin{equation}
\label{motion}
\varphi_{t}=
\left[ \begin{array}{c}
e^{i\omega t}\varphi_{0+} \\ 
e^{-i\omega t}\varphi_{0-}
\end{array}
\right],
\end{equation}
where $\omega=\frac{\mu B}{\hbar}$ and the initial state $\varphi_{0}$ is equal to
$\left[ \begin{array}{c}
\varphi_{0+} \\ 
\varphi_{0-}
\end{array}
\right]$.
Recall that the speed of evolution of the electron along $S^{3}$ was given by Eq. (\ref{speed}). Let us find the speed $\frac{ds}{dt}$ of the projection of this evolution on the space of physical states $S^{2}=CP^{1}$. For this recall that the Killing metric on $su(2)$ coincides with the Euclidean metric. The embedding of $S^{2}$ into $su(2)=R^{3}$ induces the familiar metric on $S^{2}$. Such a metric also coincides with the famous Fubini-Study metric on $CP^{1}$ identified with $S^{2}$. Since this metric is just a restriction of the Euclidean metric on $R^{3}$, one has $\left (\frac{ds}{dt}\right)^{2}=\left (\frac{dx}{dt}\right)^{2}+\left (\frac{dy}{dt}\right)^{2}+\left (\frac{dz}{dt}\right)^{2}$. Using Eqs. (\ref{xyz})-(\ref{motion}), one obtains then
\begin{equation} 
\frac{ds}{dt}=4\omega |\varphi_{0+}||\varphi_{0-}| = \frac{4\mu B}{\hbar}|\varphi_{0+}||\varphi_{0-}|.
\end{equation}
If $\theta$ is the angle between the $Z$-axis and the vector $(x,y,z)$, then $\sin^{2}\theta=1-z^{2}$ and one finds with the help of Eq. (\ref{xyz2}):
\begin{equation}
\label{projective_speed}
\frac{ds}{dt}=\frac{2\mu B}{\hbar}\sin \theta.
\end{equation}
In particular, if the initial state is an eigenstate of $\sigma_{z}$ so that $\theta=0$ or $\pi$, then the speed $\frac{ds}{dt}$ vanishes and the evolution of the electron is projectively trivial. In other words, the eigenstates of the Hamiltonian ${\widehat h}=-\mu {\widehat {\bf \sigma}} \cdot {\bf B}$ are zeros of the ``push-forward'' vector field $d\pi (h_{\varphi})$ on $CP^{1}$. Here $d\pi$ is the differential of the map $\pi$ and $h_{\varphi}=\frac{i}{\hbar}\mu {\widehat {\bf \sigma}} \cdot {\bf B}\varphi$ as before.

Recall that the integral curves $\varphi_{t}$ given by Eq. (\ref{int_curve}) are geodesics on $S^{3}$. At the same time, as one can see from Eqs. (\ref{xyz})-(\ref{motion}), the projected curves $\{\varphi_{t} \}$ are not in general geodesics in the induced metric. Moreover, there does not exist a Riemannian metric on $CP^{1}=S^{2}$ in which the projection by $\pi$ of an arbitrary geodesic on $S^{3}$ yields a geodesic on $S^{2}$. Indeed, the great circles of $S^{2}$ (parametrized by arc length) are geodesics in the Fubini-Study metric. At the same time, they are projections of geodesics on $S^{3}$. The latter fact is obvious for the equator of $S^{2}$ if one chooses $|\varphi_{0+}|=|\varphi_{0-}|$ in Eqs. (\ref{xyz})-(\ref{motion}). For any other great circle this fact is verified by a change in direction of the $Z$-axis. But the condition that the great circles are geodesics fixes the metric on $S^{2}$ up to a constant factor. This, together with the above observation that the projection $\{\varphi_{t}\}$ of a geodesic may not be a geodesic itself, proves the claim.

Suppose, as it is advocated here, that evolution by geodesics is a valid principle of quantum dynamics. Then the obtained result suggests 
that the space of states $S^{3}$ may be ``more physical'' than the projective space $CP^{1}$ of ``physical states''. In general terms, the actual, ``easy to describe'' physics may be happening on the space of states of a quantum system. However, our experiments can so far access only the projection of physical processes into the space of physical states. The projection of a process may look much less ``natural'', then the original process on the space of states.

It was verified earlier that the commutators of spin observables are related to geometry of the space of states $S^{3}$. 
%Namely, the non-commutativity of observables is due to the curvature of $S^{3}$ with radius of $S^{3}$ being equal to $\hbar$. 
Later on in the paper the geometry of a measurement process will be discussed. It is then important to know that the probability of transition from state $\varphi$ to state $\psi$ in a measurement also has ties to geometry. 
%It turns out that the probability of transition of a given state $\varphi$ to another state $\psi$ also has a clear geometric origin, established in Ref. \cite{AA}. 
%Because $P$ does not depend on the overall phase factors of the states, the underlying geometry is projective in nature.  
Namely, this probability depends only on the distance between $\{\varphi\}$ and $\{\psi\}$ in the Fubini-Study metric on the projective space of states (see Ref. \cite{AA}). 

This result is immediate in the model under consideration. In fact, assume that the state $\psi$ is the spin-up eigenstate of ${\widehat \sigma_{z}}$ (one can always assure this by a proper choice of coordinate axes). Then, according to Eq. (\ref{xyz}), one has
\begin{equation}
z=|\varphi_{1}|^2-|\varphi_{2}|^{2}.
\end{equation}
Using the normalization condition $|\varphi_{1}|^2+|\varphi_{2}|^{2}=1$, one obtains 
\begin{equation}
\label{ff}
|\varphi_{1}|^2=\frac{1-z}{2}.
\end{equation}
%and 
%\begin{equation}
%|\varphi_{2}|^2=\frac{1+z}{2}.
%\end{equation}
If $\theta \in [0, \pi]$ is the angle between the vectors representing $\{\varphi\}$ and $\{\psi\}$ on $S^{2}$, then $z=\cos\theta$ and the equation Eq. (\ref{ff}) takes the form
\begin{equation}
\label{cos}
P(\theta)=\cos^{2}\frac{\theta}{2}.
\end{equation}
Here $P(\theta)=|(\psi,\varphi)|^{2}=|\varphi_{1}|^{2}$ denotes the probability of transition. Notice that $\theta$ is the length of the arc of a great circle through $\{\varphi\}$ and $\{\psi\}$. In other words, it is a geodesic distance between the points $\{\varphi\}$ and $\{\psi\}$ on $S^{2}$ in the Fubini-Study metric. In particular, the probability $P(\theta)$ of transition between two states depends only on the distance between them. 
%The formula Eq. (\ref{cos}) holds true for {\em any} quantum system, provided the distance $\theta$ is measured in the Fubini-Study metric on the projective space of states of the system (see Ref. \cite{AA}). 

Furthermore, the uncertainty principle for the spin observables also has a simple geometrical interpretation. Since uncertainties of observables do not depend on the overall phase of a state, the geometry underlying the uncertainty principle is, once again, projective. 
The uncertainty principle for non-commuting observables ${\widehat X}, {\widehat Y}$ can be written in the form
\begin{equation}
\label{undet}
\Delta X \Delta Y \ge \frac{1}{2}\left |(\varphi, [{\widehat X},{\widehat Y}] \varphi)\right |,
\end{equation}
where $\Delta X^{2}=(\varphi, {\widehat X}^{2}\varphi)-(\varphi, {\widehat X}\varphi)^{2}$ and similarly for $\Delta Y^{2}$.
Assuming ${\widehat X}, {\widehat Y}$ are the components ${\widehat s_{x}}, {\widehat s_{y}}$ of the spin observable ${\bf {\widehat s}}=\frac{\hbar}{2}{\bf {\widehat {\sigma}}}$, one has
\begin{equation}
\label{un}
\Delta s_{x} \Delta s_{y} \ge \frac{\hbar}{2}\left |(\varphi, {\widehat s_{z}} \varphi)\right |.
\end{equation}

Let us connect this inequality with geometry of the projective space $S^{2}=CP^{1}$. In light of the geometric interpretation of the commutators of observables Eq. (\ref{normm}) and the equation Eq. (\ref{undet}), the existence of such a connection is not surprising. Indeed, from the geometry of the sphere $S^{2}$ furnished with the Fubini-Study (i.e., the usual!) metric, for any point $(x,y,z)$ on the sphere one has:
\begin{equation}
\label{un1} 
(y^{2}+z^{2})(x^{2}+z^{2})\ge z^{2}.
\end{equation}
The inequality simply says that the product of distances from a point $(x,y,z)$ on the sphere $S^{2} \subset R^{3}$ to the $X$ and $Y$-axes is at least $|z|$.
This equation is the geometric form of the uncertainty principle.
Indeed, from Eqs. (\ref{xyz})-(\ref{xyz2}) it follows that
\begin{equation}
\label{ave}
(\varphi, {\widehat \sigma}_{x}\varphi)=x, \quad (\varphi, {\widehat \sigma}_{y}\varphi)=y, \quad (\varphi, {\widehat \sigma}_{z}\varphi)=z,
\end{equation}
where $x,y,z$ are coordinates of the point of $S^{2}$ representing the state $\varphi$.
In addittion, ${\widehat \sigma}_{x}^{2}={\widehat \sigma}_{y}^{2}={\widehat \sigma}_{z}^{2}=I$ and so
\begin{equation}
\label{sxyz1}
\Delta \sigma_{x}^{2}=1-x^{2}=y^{2}+z^{2}, \quad \Delta \sigma_{y}^{2}=1-y^{2}=x^{2}+z^{2}, \quad \Delta \sigma_{z}^{2}=1-z^{2}=x^{2}+y^{2}.
\end{equation}
%We therefore see that the uncertainties of components ${\widehat s_{k}}$ of the spin operator can be written in the form
%\begin{equation}
%\label{sp_un}
%\Delta s_{k}=\frac{\hbar}{2}\left(1-x_{k}^{2}\right).
%\end{equation}
%The maximum quantum mechanical uncertainty of the spin observable ${\widehat s_{k}}$ is $\frac{\hbar}{2}$.
With the use of Eqs. (\ref{ave}), (\ref{sxyz1}), the inequality Eq. (\ref{un1}) is now equivalent to the uncertainty principle Eq. (\ref{un}).

%We therefore conclude that the non-commutativity of observables in the model and the resultant uncertainty principle for these observables find a simple explanation in the geometry of the space of states. 

%Let us investigate the geometric meaning of uncertainties of observables in more detail.
Note that using the angles $\theta_{x}, \theta_{y}, \theta_{z}$ between the coordinate axes and the vector ${\bf x}=(x,y,z)$ in $R^{3}$, one can write the equations Eqs. (\ref{sxyz1}) in the form
\begin{equation}
\Delta \sigma_{x}=\sin \theta_{x} \quad \Delta \sigma_{y}=\sin \theta_{y}, \quad \Delta \sigma_{z}=\sin \theta_{z}.
\end{equation}
In particular, the uncertainty $\Delta E$ in energy for the electron in the state $\varphi$ in the model takes the form
\begin{equation}
\Delta E=\mu B \sin\theta,
\end{equation}
where $\theta$ is the angle between the vectors ${\bf B}$ and ${\bf x}$. 
Recall that $\theta$ is a geodesic distance between the points $\{\varphi\}$ and $\{\psi\}$ on $S^{2}$. 
Therefore, the uncertainty $\Delta E$ is the largest, when $\{\varphi\}$ is furthest away from the eigenstates, i.e. when $\theta=\frac{\pi}{2}$. On the other hand, $\Delta E$ vanishes at the eigenstates of ${\widehat h}$. 

The latter statement can be generalized as follows. Let $H=C^{n}$ and let ${\widehat A}: H \longrightarrow H$ be an observable with a simple spectrum $\lambda_{1}<\lambda_{2}<...<\lambda_{n}$ and eigenfunctions $\varphi_{1}, \varphi_{2},...,\varphi_{n}$. Let $\{\varphi_{t}\}$ be a geodesic through the states $\{\varphi_{k}\}, \{\varphi_{l}\}$ in the Fubini-Study metric on the projective space of states $CP^{H}$. Then for any point $\{\varphi\}$ on the geodesic the variance $\Delta A^{2}=(\varphi, {\widehat A}^{2} \varphi)-(\varphi, {\widehat A}\varphi)^{2}$ is an increasing function of the distance from $\{\varphi\}$ to the pair of eigenstates $\{\varphi_{k}\}$, $\{\varphi_{l}\}$ in $CP^{H}$. The latter distance is simply the shortest of the distances from $\{\varphi\}$ to the states $\{\varphi_{k}\}$, $\{\varphi_{l}\}$. 

In fact, because the eigenstates are orthogonal, the probability of transition from a state $\{\varphi\}$ on the geodesic to a state $\{\varphi_{i}\}$ with $i \neq k,l$ is equal to zero. In particular, $\{\varphi\}=\{c_{k}\varphi_{k}+ c_{l}\varphi_{l}\}$ for some coefficients $c_{k}, c_{l}$. Therefore, 
\begin{equation}
\label{Delta}
\Delta A^{2}=|c_{k}|^{2}\lambda_{k}^{2}+\left(1-|c_{k}|^{2}\right ) \lambda_{l}^{2}-\left(|c_{k}|^{2}\lambda_{k}+(1-|c_{k}|^{2})\lambda_{l}\right)^{2}.
\end{equation}
If $\{\varphi\}$ coincides with $\{\varphi_{k}\}$ so that $|c_{k}|=1$, then $\Delta A$ vanishes. 
On the other hand, as $|c_{k}|$ decreases, $\Delta A$ increases until $|c_{k}|$ becomes equal to $\frac{1}{{\sqrt 2}}$. By Eq. (\ref{cos}) this means that $\Delta A$ increases with the distance from $\{\varphi\}$ to $\{\varphi_{k}\}$ until $\{\varphi\}$ becomes equally distant from  $\{\varphi_{k}\}$ and $\{\varphi_{l}\}$.

It is important to note that the ``geometric'' probability of transition formula Eq. (\ref{cos}) is valid in an arbitrary space of states $H=C^{n}$. Also, as already discussed, the relationship Eq. (\ref{normm}) between the commutators and the curvature holds true on the sphere of states $S^{2n-1}$ for observables in the subspace $V \subset u(n)$ (see the end of section 2.1). Because of that, the uncertainty principle Eq. (\ref{undet}) can be still interpreted geometrically. The infinite-dimensional case requires a different approach and needs further analysis.

%In other words, the uncertainty of an observable for the system in a state $\varphi$ is a measure of separation of $\varphi$ from the eigenstates of the observable. We will return to this observation in section 5.

\section{TENSOR PROPERTIES OF EQUATIONS IN THE MODEL}
\setcounter{equation}{0}

In the developed geometrical formulation of the model, the space of states is a manifold furnished with a Riemannian metric which defines quantum dynamics on the space. It is well known that differential geometry of manifolds admits two equivalent formulations: local coordinate and coordinate free. Let us discuss the role played by both formulations in the model.
     
Consider a pair ${\widehat X}, {\widehat Y}$ of elements of the Lie algebra $su(2)$ and the corresponding pair $X_{\varphi}={\widehat X}\varphi, Y_{\varphi}={\widehat Y}\varphi$ of the associated vector fields. By direct computation one sees that
\begin{equation}
[X_{\varphi},Y_{\varphi}]=-[{\widehat X},{\widehat Y}]\varphi,
\end{equation}
where $[X_{\varphi},Y_{\varphi}]$ is the Lie bracket of the vector fields. Recall that the integral curves of non-commuting vector fields cannot form a coordinate grid on the manifold. In particular, the integral curves of vector fields $s_{x \varphi},s_{y \varphi},s_{z \varphi}$ associated with the spin observables ${\widehat s_{x}}, {\widehat s_{y}}, {\widehat s_{z}}$ do not form a coordinate grid on $S^{3}$. By the above, these integral curves are geodesics in the Killing metric on $S^{3}$. The fact that they do not form a coordinate grid is then a direct consequence of the curvature of $S^{3}$. Instead, the fields $s_{x\varphi},s_{y\varphi},s_{z\varphi}$ form a local (non-coordinate) basis at every point of $S^{3}$.

There are, of course, many ways of choosing coordinates on $S^{3}$. One natural choice is the {\em normal} coordinate system given on a neighborhood of any point by the exponential map. If $\{e_{k}\}$ is a basis on the tangent space $T_{\varphi}S^{3}$ at $\varphi \in S^{3}$ and $A=\sum_{k}a^{k}e_{k}$ is a tangent vector, then the equation of geodesic $\varphi_{t}$ through $\varphi$ in the direction of $A$ in the normal coordinates $\varphi^{1}, \varphi^{2},\varphi^{3}$ is linear:
\begin{equation}
\varphi^{k}_{t}=a^{k}t, \quad k=1,2,3.
\end{equation}

The evolution of spin state $\varphi$ in time could be thought of as a motion along the manifold $S^{3}\times R$ with $R$ being the time axis. The direct product of the Killing metric on $SU(2)$ and the usual Euclidean metric on $R$ makes $S^{3}\times R$ into a Riemannian manifold. If $\varphi_{t}$ is a geodesic on $S^{3}$, then $(\varphi_{t},t)$ will be a geodesic on $S^{3}\times R$. Notice that for any evolution $\varphi_{t}$ of the state along $S^{3}\times R$ one can find a co-moving coordinate system on $S^{3}\times R$ in which $(\varphi_{t},t)=(\varphi_{0},t)$, i.e., the state is at rest. Such a co-moving system is directly related to the well known Heisenberg representation.

The coordinates typically used on the projective space of physical states $CP^{1}$ are homogeneous $(\varphi_{1},\varphi_{2})$ and inhomogeneous $\xi=\frac{\varphi_{2}}{\varphi_{1}}$, $\eta=\frac{\varphi_{1}}{\varphi_{2}}$ coordinates. For example, the Fubini-Study metric is usually expressed in terms of these coordinates. Other coordinate systems may be useful in applications. In particular, according to Eq. (\ref{ave}), 
the expectation values of Pauli matrices for a system in a state $\varphi$ coincide with the $x,y$ and $z$ coordinates of the point $\{\varphi\}$ in $S^{2}$. Therefore, these expectation values can be identified with local coordinates on $CP^{1}$. This fact was used in section 2 to describe the motion of state along $CP^{1}$.

%Suppose for example that $\{{\widehat U}_{k}\}, k=1,2,3$ is a set of unitary operators which are also Hermitian and such, that $\{{i\widehat U}_{k}\}$ are linearly independent elements of the Lie algebra $su(2)$.  Then, the expectation values $(\varphi, {\widehat U}_{k}\varphi)$ for any two values of $k$  can be used as (local) coordinates on $CP^{1}$. In fact, since expectation values of Hermitian operators are real, we have
%\begin{equation}
%\label{innerR}
%(\varphi, {\widehat U}_{k}\varphi)=Re(\varphi, {\widehat U}_{k}\varphi)=||\varphi|| \cdot ||{\widehat U}_{k}\varphi||cos\theta_{k},
%\end{equation}
%where $\theta_{k}$ is the angle between radius vectors  of the points $\{\varphi\}$ and $\{{\widehat U}_{k}\varphi\}$ on $S^{2}=CP^{1}$. Equation Eq. (\ref{innerR}) is an immediate consequence of the fact that the real part of an Hermitian inner product is the ordinary inner product on the realization. Now, since  $||\varphi||=||{\widehat U}_{k}\varphi||=1$, the function $\cos\theta_{k}$ with $\theta_{k} \in [0,\pi]$ is one-to-one, and the operators ${\widehat U}_{k}$ are linearly independent in $su(2)$, we conclude that any two of the parameters $x_{k}=(\varphi, {\widehat U}_{k}\varphi)$ can be used as local coordinates on the manifold $CP^{1}$ of real dimension two. In particular, the Pauli matrices are known to be unitary. Their expectation values provide the $x,y$ and $z$ coordinates of points on the sphere $S^{2} \subset R^{3}$ identified with $CP^{1}$.   

The group $SU(2)$ acting on the space $C^{2}$ is the symmetry group of the theory. In particular, the group $SU(2)$ ``extends'' to act on tensor algebra over $C^{2}$ and the Schr{\"o}dinger equation is a {\em tensor equation}. For any fixed time $t$ the Schr{\"o}dinger evolution operator ${\widehat U}(t,0)$ given by Eq. (\ref{int_curve}) is an {\em active} realization of $SU(2)$-transformations on $C^{2}$. The corresponding {\em passive} realization consists in a unitary change of basis on $C^{2}$.

One could consider instead the sphere $S^{3} \subset C^{2}$ as a base manifold for tensor bundles and make the group $SU(2)$ act locally on tensor products of spaces tangent and cotangent to $S^{3}$. For each tensor type this gives a subbundle of the corresponding tensor bundle over $C^{2}$. The Schr{\"o}dinger equation is then a tensor equation with terms which are vector fields on $S^{3}$. In fact, it is the equation for integral curves  of the vector field $h_{\varphi}=-\frac{i}{\hbar}{\widehat h}\varphi$ on $S^{3}$. Alternatively, it is the equation of geodesics in the Killing metric on $S^{3}$.

The notion of symmetry in QM is usually understood as an invariance of the Hamiltonian of the system under a symmetry transformation. In this case the Hamiltonian commutes with the transformation and the generator of transformation becomes a constant of motion. Although this is certainly true for rotations about the field direction in the model, such a restricted understanding of symmetry is not suitable for this Letter.

Mathematically, the Schr{\"o}dinger equation $h_{\varphi}=-\frac{i}{\hbar}{\widehat h}\varphi$ in the model is written in a specific basis on the space of states $C^{2}$. Under a unitary transformation ${\widehat U} \in SU(2)$ of the basis the Schr{\"o}dinger equation behaves as a vector equation. In particular, the right hand side of the equation becomes equal to $-\frac{i}{\hbar}{\widehat U}^{-1}{\widehat h}\varphi$. Since coordinates of points $\varphi$ also change to become $\psi={\widehat U}^{-1}\varphi$, the right hand side of the equation takes the form 
\begin{equation}
\label{S_T}
-\frac{i}{\hbar}{\widehat U}^{-1}{\widehat h}{\widehat U}\psi.
\end{equation}
Note that ${\widehat U}^{-1}$ and ${\widehat U}$ in Eq. (\ref{S_T}) act on different spaces! Namely, ${\widehat U}^{-1}$ acts on the tangent space $T_{\varphi}C^{2}$, while ${\widehat U}$ acts on $C^{2}$ itself. It is common, however, to identify vector spaces with the spaces tangent to them. By following this practice, one sees that under a change of basis the Hamiltonian is transformed by ${\widehat h} \longrightarrow {\widehat U}^{-1}{\widehat h}{\widehat U}$. In other words, it transforms as a tensor of rank $(1,1)$. If in addition ${\widehat h}$ and ${\widehat U}$ commute, then (and only then!) the Hamiltonian is invariant and the usual conserved quantities exist in accordance with the Noether's theorem.

A particular choice of an orthonormal basis on the space of states $C^{2}$ (alternatively, on $S^{3}$ or $CP^{1}$) has an affect on results of observations expressed in the basis. The reason for that is clear: which state is a spin-up state, for example, depends on the basis in $C^{2}$. This is quite analogous to dependency of the state of rest in classical mechanics on a choice of reference system.

Alternatively, under the identification Eq. (\ref{Mat}) of space $C^{2}$ with a space of $2\times 2$-matrices the choice of an orthonormal basis in $C^{2}$ dictates the choice of basis in the Lie algebra $su(2)$, i.e., the choice of sigma-matrices. A unitary transformation of sigma-matrices changes their eigenvectors, thus affecting the results of observations expressed in terms of these eigenvectors. 

The above can be also rephrased in terms of the geometry of the projective space $CP^{1}=S^{2} \subset R^{3}$. Namely, a unitary transformation of basis in $C^{2}$ induces an orthogonal transformation of basis on $R^{3}$.  But the choice of a basis in $R^{3}$ determines the result of measurement of the $Z$-component of electron's spin. 

One can, therefore, conclude that a choice of coordinates on manifolds $C^{2}$, $S^{3}$ and $CP^{1}$ has a physical meaning similar to the choice of a reference system on the classical space and should not be neglected.

On the other hand, physical laws must be, of course, coordinate independent. In the considered case this is assured by the geometric nature of the model.  In particular, 
the Schr{\"o}dinger equation Eq. (\ref{Schroed}) can be written in a form independent of an orthonormal basis on the space of spin states $C^{2}$, i.e., in a vector form. In this case a state is just a vector $\Phi$ of $C^{2}$ rather than a column  
$\left[ \begin{array}{c}
\varphi_{1} \\ 
\varphi_{2}
\end{array}
\right]$ 
of components in a particular orthonormal basis on $C^{2}$. The Hamiltonian is an operator ${\bf {\widehat h}}$ rather than a matrix. The equation Eq. (\ref{Schroed}) then becomes a vector equation on the manifold $S^{3}$ written in a coordinate-free form:  
\begin{equation}
\label{vector}
\frac{d \Phi_{t}}{dt}=-\frac{i}{\hbar}{\bf {\widehat h}}\Phi_{t}.
\end{equation}

Note that the eigenvalue problems for observables in the model are also tensor equations. In particular, the eigenvalue problem for the Hamiltonian can be written in a coordinate free form:
\begin{equation}
{\bf {\widehat h}}\Phi=\lambda \Phi.
\end{equation}

The tensor character of equations of quantum theory in the model signifies that the {\em principle of relativity} holds true on the space of states. This means, first of all, that both, the active transformations on the space of states and the passive transformations of coordinates on the space are available. In particular, there exist various physically distinguishable reference systems on the space of states (say, different bases on $C^{2}$). Most importantly, the validity of principle of relativity  in the theory means that the equations of the theory are the same when written in any such reference system, that is, they are {\em tensor} equations. 

One may doubt the significance of such a principle in the example. After all, the $SU(2)$ symmetry in the model has been known for years. Why would such a ``relativistic'' view of this symmetry be useful? Note however, that the quantum dynamics in the advocated approach takes place on the space of states. 
In particular, the evolution of a quantum system is the motion along a geodesic on the (curved) space of states. Because of that the notions of a reference system, of passive and active transformations, of tensor equations, as well as other differential-geometric notions on the space of states, become physically meaningful. 
This meaning is very much in line with the meaning of similar notions in special and general relativity or in the theory of gauge fields. 

For instance, a change in direction of the magnetic field ${\bf B}$ induces an active transformation on the space of states. This transformation has physically measurable consequences: it moves geodesics on the space of states and hence changes the evolution of electron's state. At the same time, the change in ${\bf B}$ can be compensated by a passive transformation. Indeed, one can choose an orthonormal basis on $C^{2}$ so that the components of ${\bf B}$ in the corresponding basis in $R^{3}$ remain the same. The equation of the new electron's path in this basis coincides then with the original one.

Most certainly, the above principle is different from the principle of relativity in space-time. In fact, it deals with tensor properties of equations on a Hilbert space of states, rather than on space-time. At the same time, it has the same kind of underlying mathematics and the same spirit as the ordinary principle of relativity.
The principle is in fact a particular instance of the {\em principle of functional relativity} introduced in Ref. \cite{Kryukov}. 

As discussed, the Hamiltonian in the model is not in general invariant under unitary transformations. In particular, an active unitary transformation associated with a change in direction of the magnetic field ${\bf B}$ produces a new Hamiltonian. In other words, the Hamiltonian is not a {\em scalar} in the tensor approach to the model. 

On the other hand, due to geometric nature of the model, one can easily identify the most important scalars (or {\em invariants}), forming the ``bone structure'' of the theory. One such invariant is the distance between any two points on spaces of states $C^{2}$, $S^{3}$ and $CP^{1}$ furnished with the above discussed metrics. Another one is the speed of quantum evolution in $S^{3}$ in a given magnetic field given by Eq. (\ref{speed}). Yet another one is the scalar curvature of $S^{3}$ or $CP^{1}$. This curvature can be expressed in terms of the sectional curvature of $S^{3}$ which was found to be $1/\hbar^{2}$.  

Let us remark that tensor character of the theory allows one to extend the original symmetry group $SU(2)$ to the group $GL(2,C)$ of general linear transformations acting on fibers of the tangent bundle over $S^{3}$. Moreover, Schr{\"o}dinger dynamics on the space of states $S^{3}$ can be formulated in a way invariant under  general coordinate transformations on $S^{3}$. This follows at once from the fact that the Schr{\"o}dinger equation in the model is the equation of geodesics on the Riemannian manifold $S^{3}$. As such, this equation is meaningful in {\em arbitrary} coordinates on the manifold.

Finally, let us comment on a possible argument against the advocated geometric approach. 
In the model considered here the Riemannian metric on the sphere $S^{3}$ has turned the sphere into a manifold of constant sectional curvature. This allowed us to relate the curvature of the metric with the Planck's constant. However, in general the sectional curvature of the Riemannian metric defined by Eq. (\ref{metric}) will not be a constant. Even when it is, there is no reason for this constant to be the same as in the model under discussion.   It seems therefore that by making the Riemannian metric depend on the Hamiltonian of the system, one shall in general loose the relationship between the curvature and the Planck's constant!

To reply to this argument recall that the model considered here is non-relativistic (in the usual sense). In particular, the equation Eq. (\ref{Pauli}) is a special case of the Pauli equation. The latter equation is well known to be the non-relativistic limit of the Dirac equation for electron in electromagnetic field. The Dirac equation can be written in the form
\begin{equation}
\label{Dirac}
i\hbar \frac{d}{dt}
\left[ \begin{array}{c}
{\widetilde \varphi} \\ 
{\widetilde \chi}
\end{array}
\right]= c {\widehat {\bf \sigma}} \cdot \left ({\widehat {\bf p}}-\frac{e}{c}{\bf A}\right)  
\left[ \begin{array}{c}
{\widetilde \chi} \\ 
{\widetilde \varphi}
\end{array}
\right]
+e\phi\left[ \begin{array}{c}
{\widetilde \varphi} \\ 
{\widetilde \chi}
\end{array}
\right]
+mc^{2}
\left[ \begin{array}{c}
{\widetilde \varphi} \\ 
-{\widetilde \chi}
\end{array}
\right],
\end{equation}
where ${\widetilde \varphi}, {\widetilde \chi}$ are two-component spinors, $(\phi, {\bf A})$ is the $4$-potential of the field, ${\widehat {\bf p}}$ is the momentum operator and $e$ is the electron's charge (see, for example, Ref. \cite{B-D}). The largest term in Eq. (\ref{Dirac}) is the one containing the $mc^{2}$ factor. By substituting  
\begin{equation}
\label{non-relat}
\left[ \begin{array}{c}
{\widetilde \varphi} \\ 
{\widetilde \chi}
\end{array}
\right]=
e^{-\frac{imc^{2}}{\hbar}t}
\left[ \begin{array}{c}
\varphi \\ 
\chi
\end{array}
\right]
\end{equation}
into Eq. (\ref{Dirac}) one recovers in the standard way the Pauli equation for the spin state $\varphi$ with values in $C^{2}$.

Assume that the $4$-potential in Eq. (\ref{Dirac}) describes a weak homogeneous magnetic field ${\bf B}$ and let ${\widehat h}_{D}$ be the corresponding Hamiltonian. Consider the metric $G_{D}$ given by Eq. (\ref{metric}) with the Hamiltonian ${\widehat h}_{D}$. Then solutions to the Dirac equation for electron in the field ${\bf B}$ are geodesics in this metric. Note that the Hamiltonian ${\widehat h}_{D}$ has in the non-relativistic limit the form $mc^{2}+{\widehat h}$, where ${\widehat h}$ is the non-relativistic Hamiltonian used in Eq. (\ref{Pauli}). Accordingly, the metric $G_{D}$ can be written in the form $(\hbar/mc^{2})^{2} ( I+\epsilon)$, where $\epsilon$ is a small correction due to the Hamiltonian ${\widehat h}$ and $I$ is the identity. It follows that the sectional curvature of $G_{D}$ consists of the main term of the order $(mc^{2}/\hbar)^{2}$ and a small correction due to ${\widehat h}$. Since the main term is constant (i.e., it does not depend on the fields), the advocated geometric interpretation of the curvature remains possible. 
However, a more careful analysis of the situation requires a ``functional relativistic'' formulation of the problem and will be discussed elsewhere.

\section{THE PROCESS OF MEASUREMENT}
\setcounter{equation}{0}

In the model under consideration the Schr{\"o}dinger equation is the equation of geodesics on the space of states $S^{3}$ furnished with the Killing metric. That means that the dynamics in the theory takes place on the Hilbert space of states rather than on the classical space. In this and the following sections it will be argued that the space of states is also the most appropriate background for tackling problems related to quantum measurement. In particular, the process of collapse of a state can be regarded as a geodesic motion in the space of states with the metric ``skewed'' by the measuring device.

Consider a pair of spin-$1/2$ particles.  In QM the most general spin state of such a pair has a form
\begin{equation}
\label{sss}
\Psi=\sum_{i,j=\pm}c_{ij}\varphi_{i}\psi_{j},
\end{equation}
where $\varphi_{+}, \psi_{+}$ are spin-up, and $\varphi_{-}, \psi_{-}$ spin-down states of the particles. The Hilbert space of states having such a form is the tensor product $C^{2} \otimes C^{2}$. The unit normalized states form a sphere $S^{7}$ in this four-dimensional complex space. 

Whenever $\Psi$ is not a product of states of the particles, the state of the pair is called {\em entangled}. It is well known that, when the particles are microscopic (i.e., sufficiently small in mass and size), the entangled states do indeed exist. 
By assuming the universal validity of QM, one concludes that the entangled states can be also prepared when one of the particles is replaced with a macroscopic measuring device, designed to measure spin of the second particle. In this case the total state of the system has the form $\Psi=a\varphi_{+}\psi_{+}+b\varphi_{-}\psi_{-}$ where $\psi_{\pm}$ represent states of the device, corresponding to spin-up and spin-down outcomes of measurement.
However, unlike the case of microscopic objects, the entangled states with macroscopic objects have never been observed in experiments. 
%Even the ordinary non-trivial superpositions of position eigenstates of a macroscopic object have never been observed. QM seems to make wrong predictions about macroscopic objects.  
The phenomenon of decoherence does not help resolve this problem because the mixtures of states of macroscopic objects have not been observed in experiments either. 

Recall that in the classical physics the motion of a pair of interacting particles on a manifold can be thought of as a motion of a point in a higher dimensional configuration space. Suppose in particular that particles of masses $m$ and $M$ interact gravitationally and move in the space $R^{3}$ in accordance with the Newton's Second Law. Then the motion of the pair is represented by a trajectory in  configuration space $R^{6}$. However, if $M \gg m$, the motion simplifies and can be thought of as a motion of the particle of mass $m$ in the field created by the particle of mass $M$. In this case the configuration space $R^{6}$ of the pair is ``effectively reduced'' to the space $R^{3}$ and the field on $R^{3}$ created by the heavier particle.

An analogous ``reduction'' of configuration space is implicitly present in the unitary QM whenever the influence of a ``macroscopic surrounding'' of a quantum system is accounted for by an appropriate choice of potential in the Schr{\"o}dinger equation. 

Let us explore the idea that a similar approach can be applied to the process of measurement in the model. Namely, assume that
the motion of the total state function of the electron and the measuring device during their interaction can be effectively replaced with the motion of electron's state function in $S^{3}$ under the influence of a {\em physical field} on $S^{3}$ created by the measuring device. This means that, in some sense, the state function of the device does not change much as a result of interaction. 

One immediate objection is that the observed states of the device are orthogonal and so the state cannot change ``just a little''. Without addressing this problem in detail, let us point out that the metric on the space of states of the device may be ``skewed''. As a result, two different position eigenstates of a pointer may become very close in this metric. 
For instance, let $H$ be the Hilbert space obtained by completion of a space of ordinary functions on $R^{3}$ with respect to the inner product
\begin{equation}
\label{mett}
(\varphi,\psi)_{H}=\int e^{-({\bf x}-{\bf y})^{2}}\varphi({\bf x})\psi({\bf y})d^{3}{\bf x}d^{3}{\bf y}.
\end{equation}
Such a space contains in general the eigenstates of position operator, i.e., the delta-functions $\delta({\bf x}-{\bf a})$. Moreover, two different position eigenstates $\delta({\bf x}-{\bf a})$, $\delta({\bf x}-{\bf b})$ with $\left\|{\bf a}-{\bf b}\right\|_{R^{3}} \ll 1$ are close in the metric Eq. (\ref{mett})
(see Ref. \cite{Kryukov}). If, in particular, $\delta({\bf x}-{\bf a})$, $\delta({\bf x}-{\bf b})$ are the eigenstates of a pointer, then the fact that theses states are close in $H$ can imply that the state of the pointer does not change much in the process of interaction with a measured microscopic system.

This argument does not prove, of course, that the proposed ``reduction formalism'' can be consistently implemented into the theory. To validate the formalism one must demonstrate that all imaginable measurements in QM can be modeled ({\em at least} in terms of their outcomes) by a field on the Hilbert space of states of the measured system. In what follows such a demonstration will be presented in the case of a finite dimensional space of states and a time-independent observable. Namely, in this case a specific working model of measurement based on a perturbation of the metric on the space of states of the system will be constructed. Moreover, the effectiveness  and the scope of the proposed method suggest that it can be successfully applied in general.

For guidance in modeling the process of measurement let us return to the example under consideration. Suppose that the device in the example measures the component of electron's spin in the direction of magnetic field. Equivalently, since the Hamiltonian is given by ${\widehat h}=-\mu {\bf {\widehat \sigma}} \cdot {\bf B}$, the device can be designed to measure the electron's energy. 
Without loss of generality one may assume that the field is directed along the $Z$-axis. Then the two eigenstates $\{\psi_{1}\}, \{\psi_{2}\}$ of ${\widehat h}$ in $CP^{1}$ are positioned on the $Z$-axis at the poles of the sphere $S^{2}=CP^{1}$. Recall that these eigenstates are zeros of the vector field $h_{\varphi}=-\frac{i}{\hbar}{\widehat h} \varphi$ projected on $CP^{1}$. In particular, the Schr{\"o}dinger evolution of $\psi_{1}, \psi_{2}$ is projectively trivial. So, as a result of the interaction between the electron and the device, the original circular motion of the electron's state along a parallel on $S^{2}=CP^{1}$ is changed to the state of rest at one of the poles on the sphere. 
The following hypothesis, which will be clarified and exemplified later on, seems to be in order:  

\begin{description}
\item[(H1)]
{\em The measuring device creates a physical field on the sphere of states with sources at the eigenstates of the measured observable. This field is capable of driving the electron's state toward one of the eigenstates.}
\end{description}

%%%%%%%%%%%%%%%%%%%%%%%%%%%%%%%%%%%%%%%%%%%%%%%%%%%%%%%%%%%%%Move the block below
%%%%%%%%%%%%%%%%%%%%%%%%%%%%%%%%%%%%%%%%%%%%%%%%%%%%%%%%%%%%%%%

Note that in the position measuring experiments a measuring device is a system of counters distributed in space at the eigenstates of the position observable. The counters are indeed sources of interaction between the particle and the apparatus. The points in the classical space where the measured particle can be found, can be identified with the position eigenstates of the particle (see Refs. \cite{Kryukov}, \cite{Kryukov3}). Since the counters are also located at these points, one can identify them with sources in the space of states of the particle. 
Finally, because the sources are capable of catching the particles, the (loosely stated) hypothesis {\bf (H1)} is satisfied. This gives one the hope that a specific form of the hypothesis can be, in fact, realized.

%When measuring an electron's spin by a Stern-Gerlach apparatus, we actually measure, once again, the electron's position. However, as a result of passing through the apparatus, the position and the spin eigenstates of the electron become entangled. Because of that, the interaction driving electron's state to a particular position, also drives it to a particular spin eigenstate. The hypothesis {\bf (H1)} suggests in this case that the interaction ``extends'' to the space of spin states and is centered there at spin eigenstates.
%%%%%%%%%%%%%%%%%%%%%%%%%%%%%%%%%%%%%%
%%%%%%%%%%%%%%%%%%%%%%%%%%%%%%%%%%%%%Move the above block

What could be the nature of the field postulated in the hypothesis? Recall that the Riemannian metric in the model is dynamical, i.e., it drives the evolution of the electron's state. Suppose that the field is nothing but a perturbation of the metric on the sphere of states $S^{3}$, induced by the presence of the measuring device. 
The evolution of the electron's state during the measurement is then the motion along a geodesic in the perturbed metric. Note that the metric on the total space $S^{3}$ and not only on the base space $CP^{1}=S^{2}$ must be perturbed. In fact, as already discussed, the projection $\{\varphi_{t}\}$ of a Schr{\"o}dinger evolution $\varphi_{t}$ is not in general a geodesic. 

The above approach is certainly attractive, in particular, it does not require any ad hoc features in the theory. Furthermore, the approach can be easily implemented by an appropriate ``denting'' of the sphere of states of the system. 
Namely, as shown below, by perturbing the metric on the sphere $S^{3}$ one can ``redirect'' the evolution of electron's state so that the state would become stationary. 

For a greater generality, assume that the Hilbert space $H$ of states of the system is $n$-dimensional and let $(\varphi^{1},...,\varphi^{n})$ be the usual coordinates on $H=C^{n}$.  Let the Riemannian metric on $H$ to be of the form
\begin{equation}
\label{eta_delta}
g_{ik}=\eta^{2} \delta_{ik},
\end{equation}
where $\eta=\eta(\varphi)$ is a function on $H$. 
%
%The length functional for this metric is
%\begin{equation}
%l[\varphi_{t}]=\int^{t_{2}}_{t_{1}} \sqrt {\eta^{2} (\varphi)\delta_{ik}\frac{d \varphi^{i}}{dt}\frac{d \varphi^{k}}{dt}}dt.
%\end{equation}
%
The equation of geodesics in this metric can be obtained in the usual way by variation of the length functional on paths $\varphi_{t}$. Let $s$ be the arc length parameter and let $\tau$ be a parameter defined by $d\tau=\frac{ds}{\eta}$. Then the equation of geodesics in the metric Eq. (\ref{eta_delta}) on $H$ can be written in the form
\begin{equation}
\label{ray}
\frac{d^{2} \varphi_{\tau}}{d\tau^{2}}=\frac{1}{2}\nabla \eta^{2},
\end{equation}
where $\varphi_{\tau}$ is identified with $\varphi_{t(\tau)}$. 
A similar equation in $R^{3}$ is well known in geometrical optics where it describes propagation of rays in a media with refractive index $\eta$. The equation Eq. (\ref{ray}) is also similar to the Newton equation of motion for a unit mass in the field $U=-\eta^{2}/2$. 

The form of equation Eq. (\ref{ray}) makes it easy to see that for any sufficiently smooth path $\varphi_{\tau}$ there exists a function $\eta$ such that the equation is satisfied, at least on a neighborhood of $\varphi_{0}=\left. \varphi_{\tau}\right|_{\tau=0}$. That is, $\varphi_{\tau}$ is a geodesic through $\varphi_{0}$ in the Riemannian metric Eq. (\ref{eta_delta}) on $H$, at least for small values of $\tau>0$. That also means that an arbitrary sufficiently smooth path $\varphi_{\tau}$ with values on the sphere $S^{H} \subset H$ is a geodesic through $\varphi_{0}=\left. \varphi_{\tau}\right|_{\tau=0}$ in an appropriate Riemannian metric on $S^{H}$, at least for small $\tau>0$. As a side remark, various global results of this kind can be obtained by applying the methods of geometrical optics to Eq. (\ref{ray}) (see Ref. \cite{Krav} for a review of geometrical optics).

Consider now an arbitrary non-stationary Schr{\"o}dinger evolution $\varphi_{t}$ on $S^{H}$ driven by an invertible time-independent Hamiltonian ${\widehat h}$. Let $S^{H}$ be furnished with a Riemannian metric $G_{R\varphi}$ in which the evolution is a motion along a geodesic. It is known that such a Riemannian metric on $S^{H}$ exists.  Pick a moment of time $t=a$ and an eigenstate $\psi \in S^{H}$ of the Hamiltonian ${\widehat h}$. Consider the geodesic $\chi_{t}$ connecting $\varphi_{a}$ and $\psi$. By Eq. (\ref{ray}), there exists a perturbation of the metric $G_{R\varphi}$ on a small neighborhood of $\varphi_{a}$ which transforms the geodesic $\varphi_{t}$ into the geodesic $\chi_{t}$. Note that one could similarly perturb the metric on a small neighborhood of $\psi$ to transform $\chi_{t}$ into the stationary geodesics through $\psi$. 

The provided perturbation of the metric takes place on the sphere of state $S^{H}=S^{2n-1}$. What could one say about the metric and the motion on the projective space $CP^{n-1}=S^{2n-1}/S^{1}$? 
To answer, consider the Riemannian metric on $C^{n}_{\ast}$ (i.e., $C^{n}$ without the origin), defined for all $\varphi \in C^{n}_{\ast}$ and all pairs of vectors $\xi, \eta \in T_{\varphi}C^{n}_{\ast}$ by 
\begin{equation}
\label{newMetric}
G_{R\varphi}(\xi, \eta)=\frac{Re\left({\widehat h}^{-2}\xi, \eta \right)_{C^{n}}}{||\varphi||^{2}_{C^{n}}}.
\end{equation}
This metric, being restricted to the sphere $S^{2n-1}$, is a particular case of the metric Eq. (\ref{metric}). Solutions to the Schr{\"o}dinger equation are geodesics in the metric Eq. (\ref{newMetric}) on $C^{n}_{\ast}$ (and in the induced metric on the sphere, see Ref. \cite{Kryukov}).
%
%Be careful here as there are (anti)Hermitian operators on C^{n} which are not in su(n)!!!
%
%Notice that this metric being restricted to the sphere $S^{2n-1}$ coincides with the ordinary metric on $S^{2n-1}$ induced by embedding of the sphere into the Euclidean space $C^{n}$. As already mentioned, such a metric can be also obtained by identifying the sphere $S^{2n-1}$ with the quotient $SU(n)/SU(n-1)$ of unitary groups and inducing the metric on the sphere from the Killing metric on $SU(n)$ via the embedding of $S^{2n-1}$ into $SU(n)$.
%MAKE SURE TO CORRECT!!!

Notice that the multiplication map $\lambda: \varphi \longrightarrow \lambda \varphi$ with $\lambda \in C_{\ast}$ is an isometry of the metric Eq. (\ref{newMetric}), that is, $G_{R\lambda \varphi}(d\lambda \xi,d\lambda \eta)=G_{R\varphi}(\xi, \eta)$. This is clear because $d\lambda=\lambda$ by linearity of the map and $Re(\lambda \xi,\lambda \eta)_{C^{n}}/||\lambda\varphi||^{2}_{C^{n}}=Re(\xi,\eta)_{C^{n}}/||\varphi||^{2}_{C^{n}}$. Because of that, the metric Eq. (\ref{newMetric}) ``projects down'' to $CP^{n}$, giving a metric on the projective space. More precisely, the metric Eq. (\ref{newMetric}) is induced by the projection of $C^{n}_{\ast}$ onto $CP^{n-1}$, furnished with a Riemannian metric. Provided ${\widehat h}^{2}$ is proportional to the identity operator, the latter metric coincides with the Fubini-Studi metric on $CP^{n-1}$ (see Ref. \cite{Kryukov}). 

%The question now is whether the perturbation of the metric on $S^{2n-1}$, needed to ``redirect'' the Schr{\"o}dinger evolution to account for the process of measurement, could preserve this property. 

Note however that the multiplication by a complex number may not remain an isometry of the perturbed metric on $C^{n}_{\ast}$. That is, the metric on $C^{n}_{\ast}$, needed to ``redirect'' the Schr{\"o}dinger evolution to account for the process of measurement, may not originate in a Riemannian metric on $CP^{n-1}$. In mathematical terms, the projection of $C^{n}_{\ast}$ with a perturbed metric onto $CP^{n-1}$ is not in general a Riemannian submersion. In particular, such is the case for the discussed local perturbation of the metric on $C^{n}_{\ast}$. This suggests once again that the metric on the sphere $S^{H}$ in the approach under investigation has a greater significance than the metric on the projective space $CP^{H}$.

%%%%%%%%%%%%%%%%%%%%%%%%%%%%%%%%%%%%
This analysis demonstrates that by an appropriate ``denting'' of the sphere of states $S^{H}$, $\dim H < \infty$, one can locally affect geodesics on the sphere in a desirable fashion. In particular, by perturbing the Riemannian metric $G_{R\varphi}$ on $S^{H}$ one can alter the Schr{\"o}dinger evolution of the state and drive the state toward one of the eigenstates of the measured observable. It follows that the physical field in the hypothesis {\bf (H1)} can be indeed identified with a perturbation of the metric $G_{R\varphi}$. Note that the resulting metric is time-independent. The electron's state in the construction propagates along a geodesic on the sphere of states, runs into a region with perturbed metric and collapses.

%Recall now that the equation of geodesics in the metric Eq. (\ref{metric}) on $S^{H}$ was just the Schr{\"o}dinger equation. At the same time, the equation of geodesics Eq. (\ref{ray}) does not coincide in general with the Schr{\"o}dinger equation. That is, the change in the metric needed to drive the state to an eigenstate results in general in a non-Schr{\"o}dinger evolution of the state.
It is important to remark that the above demonstration is only an {\em existence} proof; it does not provide a realistic model of interaction between the system and the device. Moreover, a particular nature of the field in the hypothesis {\bf (H1)} will not be essential in the following.  The thorough analysis of this nature requires the equations of the field and is left for the upcoming publications. The mere existence of the field satisfying the needed properties will be sufficient for the purpose of this Letter. 

%is by no means physical and it was not intended as such. The goal here was to make an {\em existence} statement, rather than to create a realistic model.
%A particular nature of the field postulated in the hypothesis {\bf (H1)} will not be essential in the following.  The thorough analysis of this nature requires the equations of the field and is left for the upcoming publications. The mere existence of the field satisfying the needed properties will be sufficient for the purpose of this letter. 

%Moreover, the assumption is easy to implement mathematically by simply ``denting'' the sphere $S^{3}$ at the eigenstates. Such a deformation of the sphere will affect geodesics on a neighborhood of eigenstates and the resulting evolution of the state along $S^{3}$ and $CP^{1}$. However, the assumption will not be essential in the following and its analysis is left for the upcoming publications. 

The postulated physical field may be able to drive the state to one of the eigenstates, i.e., it may be responsible for the collapse  {\em itself}. However, in such a scenario collapse seems to be a deterministic process and the probabilistic nature of collapse to a particular eigenstate is not explained. Recall now that in accordance with Eq. (\ref{cos}), the probability of collapse of a given state $\varphi$ to an eigenstate $\psi_{k}$ of an observable depends only on the distance $\theta$ between the states in the Fubini-Study metric on the projective space of states. This crucial property allows one to resolve the remaining difficulty in creating a working probabilistic model of collapse. 
%based on perturbation of the metric on the space of states. 

Indeed, suppose that the field sources in the hypothesis {\bf (H1)} are not fixed at the eigenstates $\psi_{k}$ but fluctuate randomly about the eigenstates. In particular, the projections of the sources fluctuate randomly about the points $\{\psi_{k}\}$ on $CP^{H}$. Suppose further that fluctuations with projections of a small (in the Fubini-Study metric) amplitude are more likely to occur. Suppose finally that if a source reaches a small neighborhood of the state $\varphi$, it alters the evolution of the state and diverts it to the corresponding eigenstate (say, by perturbing the metric on the neighborhood). Then, the closer the state $\{\varphi\}$ is to a particular eigenstate $\{\psi_{k}\}$ (consequently, the larger the modulus of the coefficient $c_{k}$ in the decomposition $\varphi=\sum_{i} c_{i}\psi_{i}$ is), the more likely it becomes for the source fluctuating about $\psi_{k}$ to reach (and collapse) the state. At the same time, the further $\{\varphi\}$ and $\{\psi_{m}\}$ are (and hence, the smaller the modulus of $c_{m}$ is), the less likely it becomes for the source fluctuating about $\psi_{m}$ to reach the state. In such a way, the competition between the sources can lead to the standard Born rule for the probability of collapse. 

%Before analyzing this scenario and comparing it with the existing models of collapse, 
To prove the latter claim, let us first of all make the above assumptions precise. To keep the analysis simple, the assumptions will refer to the fibre bundle $\pi: S^{3} \longrightarrow S^{2}$ corresponding to the model under discussion. However, it will be clear that the measuring process on {\em any} fibre bundle $\pi: S^{2n-1} \longrightarrow CP^{n-1}$ and for {\em any} time-independent observable on the corresponding space of states can be treated in the same way. 

Let $\theta \in (-\pi, \pi]$ and $\alpha \in (-\pi/2, \pi/2]$ be the angular coordinates on the projective space $CP^{1}=S^{2}$. Here the coordinate curves $\alpha=\alpha_{0}$ yield great circles (pairs of meridians) through the poles $\theta=0, \theta=\pi$ of $S^{2}$ and the curves $\theta=\theta_{0}$ yield half the parallels on $S^{2}$. As before, the eigenstates $\{\psi_{1}\}$, $\{\psi_{2}\}$ are located at the poles of $S^{2}$. Let $\beta \in (-\pi,\pi]$ be the phase of a state on the sphere $S^{3}$. Then the triple $(\theta, \alpha, \beta)$ form a coordinate system on $S^{3}$. In terms of these coordinates the following hypothesis is now accepted: 

\begin{description}
\item[(H2a)]
{\em Fluctuations of each source along the sphere of states $S^{3}$ can be described by a three-dimensional stochastic process $(\theta_{t}, \alpha_{t}, \beta_{t})$. For instance, consider the source associated with the eigenstate at $\theta=0$. For any $t=t_{0}$ the random variables $\theta_{t_{0}}, \alpha_{t_{0}}, \beta_{t_{0}}$ describing the source are independent. For any $t=t_{0}$ the probability density of the random variable $\theta_{t_{0}} \in (-\pi, \pi]$ is equal to $\frac{1}{\pi}cos^{2}\frac{\theta}{2}$. The random variables $\alpha_{t_{0}}, \beta_{t_{0}}$ are uniformly distributed. The mean function of each process is zero. For any two times $t_{1}, t_{2}$, $t_{1}\neq t_{2}$, the random variables $\theta_{t_{1}}, \theta_{t_{2}}$ are practically statistically independent, so that the stochastic process $\theta_{t}$ is uncorrelated in time. In other words, $\theta_{t}$ is ideally a {\em white noise} process. The same is true about the processes $\alpha_{t}, \beta_{t}$. The stochastic processes describing different sources are independent. }

\item[(H2b)]
{\em A source at a point $\varphi \in S^{3}$ with coordinates $(\theta, \alpha, \beta)$ may be identified with a perturbation of the metric on a small neighborhood $U \subset S^{3}$ of the point. If at some time $t$ the $U$-neighborhood of a particular source contains the electron's state, the perturbation of the metric alters the evolution of the state and collapses it to the corresponding eigenstate. }
%(see Figure 1).
\end{description} 

Is there a realization of the hypothesis? It was already verified that {\bf (H2b)} can be realized for any Schr{\"o}dinger evolution $\varphi_{t}$ by a ``lensing'' effect, i.e., by redirecting $\varphi_{t}$ toward the eigenstate. 
Also, the white noise process postulated in {\bf (H2a)} certainly exists as a mathematical idealization. Moreover, the processes of this kind are common in physics. Probably the most appropriate example is the {\em thermal noise}, i.e., 
the random process describing the electric current created by the thermal motion of electrons inside a conductor. 

Could the random fluctuations of the sources in {\bf (H2a)} be of a similar origin? During a measurement the measuring device interacts with the measured system. At the same  time, the molecules (atoms, particles) of the device experience a random thermal motion. In a general (non-stationary) case fluctuations of molecules result in fluctuations of their states on the space of states. So the main new assumption made is that the interaction between the measured system and the device also takes place {\em on the space of states} of the system rather than on the classical space alone.
Fluctuations of states of the molecules are then associated with fluctuations of the field sources along the space of states leading to the postulated stochastic process. 

What is the probability $dP_{1}$ of collapse of a state $\varphi_{t}$ to a particular eigenstate $\psi_{1}$ at some specific time $t=t_{0}$ in the hypothesis? Such a probability is equal to the probability for the $U$-neighborhood of the corresponding source to contain the state at this time. Let $(\theta_{0}, \alpha_{0}, \beta_{0})$ be coordinates of $\varphi_{t_{0}}$ in the chosen coordinate system on $S^{3}$. If $U$ is sufficiently small, the change in the probability density of $\theta_{t_{0}}$ across $U$ can be neglected, and, therefore,
\begin{equation}
\label{collProb}
dP_{1}=\frac{1}{2\pi^{3}} \cos^{2}\frac{\theta_{0}}{2}dV.  
\end{equation}
Here $dV$ is the volume of $U$ which in this simplest case is identified with $d\theta d\alpha d\beta$ for some fixed values of the differentials. \footnote{The volume element for the sphere $S^{3}$ with the usual metric in the chosen coordinates is $d_{S}V=\frac{1}{2}\sin\frac{\theta}{2}\cos\frac{\theta}{2}d\theta d\alpha d\beta$. It would be more appropriate to associate $d_{S}V$ with the volume of $U$. Moreover, the expression $\sin\frac{\theta}{2}\cos\frac{\theta}{2}$ is the derivative of $\cos^{2}\frac{\theta}{2}$. This leads one to interesting models in which at any $t=t_{0}$ the random variable $\theta_{t_{0}}$ is uniformly distributed on $(-\pi, \pi]$ and the coefficient $\cos^{2}\frac{\theta}{2}$ appears in the (complementary) cumulative distribution function due to the factor $\sin\frac{\theta}{2}\cos\frac{\theta}{2}$ in the volume element $d_{S}V$. However, the element $dV$ will be sufficient for the purpose of this Letter.  } 
According to the hypothesis, the random variables describing the positions of different sources at $t=t_{0}$ are independent. Therefore, the probability  $dP_{2}$ of collapse of the state $\varphi_{t_{0}}$ to another eigenstate $\psi_{2}$ can be computed in the same way. Finally, since the stochastic processes describing the sources are uncorrelated in time, the probability for the $U$-neighborhood of a source to contain any particular point $\varphi_{0}$ is not affected by the previous history of the source. In particular, the probability rule Eq. (\ref{collProb}) is universally valid. 

On the other hand, according to Eq. (\ref{cos}), the expression 
\begin{equation}
\label{c1}
|c_{1}|^{2} =\cos^{2}\frac{\theta_{0}}{2}
\end{equation}
represents the standard probability of transition from the state $\{\varphi_{t_{0}}\}$ to the state $\{\psi_{1}\}$, provided $\theta_{0}$ is the distance between the states in the Fubini-Study metric and $c_{1}$ is the coefficient of $\psi_{1}$ in the decomposition of $\varphi_{t_{0}}$.  Clearly, the distance $\theta_{0}$ in the formula Eq. (\ref{c1}) can be replaced with the the angle $\theta_{0}$ between the states, explaining the chosen notation. Of course, a similar formula holds true for the coefficient $c_{2}$ of $\psi_{2}$. It follows that the ratio $dP_{1}/dP_{2}$ coincides with $|c_{1}|^{2}/|c_{2}|^{2}$. The conclusion is that the postulated hypothesis yields the Born rule for the probability of collapse as was claimed.

The ``single-push'' process of collapse of the state $\varphi_{t}$ to an eigenstate can be replaced with a more elaborate stochastic process. Each encounter with a source in this process results in a decrease in the distance $\theta \in [0,\pi]$ between the state $\{\varphi_{t}\}$ and the corresponding eigenstate $\{\psi_{k}\}$ by a certain value $\delta$. Between the encounters the state undergoes the ordinary Schr{\"o}dinger evolution. Assume for simplicity that the frequency of encounters is sufficiently high. In this case one can neglect the Schr{\"o}dinger evolution of the state during the measurement. The stochastic process of collapse can be then defined as a finite, time-homogeneous Markov chain with absorbing boundaries $\theta=0$ and $\theta=\pi$ and with the number of states equal to $\pi/\delta+1$. The transition matrix for the process can be found via simple formulas from the condition that the steady-state transition matrix has the right transition probabilities from any state $\theta_{i}$ to the absorbing states $\{\psi_{1}\}, \{\psi_{2}\}$ (namely, $\cos^{2}\frac{\theta_{i}}{2}$ and $\cos^{2}\frac{\pi-\theta_{i}}{2}=\sin^{2}\frac{\theta_{i}}{2}$). The resulting process is a generalization of the (biased) random walk with absorbing boundaries (also known as the gambler's ruin), in which the transition probabilities vary with the state. Namely, the transition probabilities for a step toward an absorbing state $\{\psi_{i}\}$ increase as the electron's state moves closer to $\{\psi_{i}\}$.

Various stochastic processes have been extensively used in modeling collapse (see reviews Refs. \cite{Pearle} and \cite{Bassi}). In general words, the existing models are based on adding an external random noise term and a term containing the measured observable to the Schr{\"o}dinger equation. The term with the observable provides the ``choices'' for observations, while the random noise term is a ``chooser'' (see Ref. \cite{Pearle}).  The probability density for a particular noise in the models is given by yet another equation. This equation makes it more probable for the noise to fluctuate around values associated with the eigenstates of the observable and in such a way that the probability of the noise also depends on the initial state $\varphi_{0}$ of the system.
When applied to the process of measurement, the models of this kind explain the probabilistic results of observations by relating them to the random noise, selected by the mentioned probability rule.
At the same time, the physical reason for a particular random noise remains unexplained (see Ref. \cite{Pearle}). 

Even without analyzing the existing stochastic models of collapse in detail, one can pinpoint the essential difference of the model considered here. Namely, the noise in the advocated approach is a process on the space of states which does {\em not} depend on a particular state $\varphi$ of the measured system. In particular, the noise in the model does not change when the state $\varphi$ changes. 
%For instance, the probability density function of the random variable $\theta_{t_{0}}$ stays the same no matter what the state $\varphi$ is. 
This independence of the noise from the state of the measured system opens a way for associating it with the measuring device itself. For example, as already discussed, the noise may originate in the thermal motion of molecules of the device, considered as a process on the space of states.

Another important observation is that the process of collapse in the model is a {\em deterministic} process on the space of states. In fact, by associating the random noise with a physical process, one should be able, in principle, to provide a specific functional form of the noise. In this case it becomes possible to predict the time and the outcome of collapse for an arbitrary evolution $\varphi_{t}$ of the system. 

%There is yet another extremely interesting observation. Let $\varphi=(\theta,\alpha,\beta)$ be a point on $S^{3}$ and let $U$ be a small neighborhood of $\varphi$ with Lebesgue measure $dV=d\theta d\alpha d\beta$ . Recall that according to Eq. (4.6), the probability $dP$ that at any time $t$ the source, described by the stochastic process $(\theta_{t},\alpha_{t},\beta_{t})$ will be found within the neighborhood $U$ is proportional to $\cos^{2}\frac{\theta}{2}d\theta d\alpha d\beta$. Let us introduce a new angular coordinate $\gamma$ by $\gamma=\frac{\pi-\theta}{2}$. In terms of $\gamma$ the probability $dP$ is proportional to $\sin^{2}\gamma d\gamma d\alpha d\beta$. The coordinate $\gamma$ has a simple geometric meaning that can be described as follows. Let the sphere $S^{2}$ identified with $CP^{1}$ be embedded into $C^{2}$ in such a way that the poles $\theta=0, \theta=\pi$ have Cartesian coordinates $(0,0)$ and $(1,0)$ respectively. Let the sphere of states $S^{3}$ be the sphere of radius $1$ centered at the origin in the same space $C^{2}$. Then $\gamma$ is the angle between the axis through $(1,0)$ and the line connecting the origin with $\varphi$ in $C^{2}$. In other words, the expression $\sin^{2}\gamma d\gamma d\alpha d \beta$ is the ordinary volume element of the sphere $S^{3}$. 

Note that the proposed mechanism of collapse, although particularly simple, is not the only one satisfying the hypotheses {\bf (H1)}, {\bf (H2)}. Furthermore, as already mentioned, the proposed mechanism is far from being realistic at this stage.
The ultimate choice of a physically valid scenario of collapse depends crucially on the field equations on the space of states and cannot be provided at this time. Instead, let us  demonstrate that 
under the above assumptions even such a simple mechanism sheds new light on the quantum measurement problem.

\section{THE MEASUREMENT PROBLEM}
\setcounter{equation}{0}

%Finally, it is important to note that the device is not left out of the picture...
%To proceed, let us list the most important observations made so far in the Letter and in Ref. \cite{Kryukov}:
%The following results obtained in Ref. \cite{Kryukov} and in this Letter can be used to support the above statements: 

%Let us summarize the main observations made so far. First of all, the Schr{\"o}dinger evolution of an arbitrary quantum system governed by a time-independent, invertible Hamiltonian can be identified with a geodesic motion on the space of states $H$ of the system furnished with an appropriate Riemannian metric. 
%As shown in Ref. \cite{Kryukov}, this result is true in the infinite-dimensional case as well. 

%Next, assume that the space of states is finite dimensional, so that $H=C^{n}$ and let $S^{2n-1}$ be the sphere of states in $H$. Then, by identifying $S^{2n-1}$ with the homogeneous space $U(n)/U(n-1)$ and by using the Killing metric on the unitary group $U(n)$ one obtains a simple geometric interpretation of the Planck's constant and the commutators of observables in terms of the curvature of the sphere $S^{2n-1}$.  

%Finally, if $H=C^{n}$, the effect of the measuring device on the measured system can be modeled by a perturbation of the metric on the sphere of states $S^{2n-1}$ of the system. The process of measurement of an arbitrary time-independent observable can be modeled by random fluctuations of the perturbation of the metric on the sphere.

The observations made so far, combined with the results of Ref. \cite{Kryukov} suggest the following statements about objects and interactions in QM:  
\begin{description}
	\item [(S1)] 
Physical objects in QM are most adequately represented by points of a Hilbert manifold of states. In this sense, they have a {\em functional} nature. 
	\item [(S2)]	
Physical interactions involving microscopic objects (in particular, the process of measurement) are most adequately described as processes on the manifold of states, rather than on the classical space alone. In other words, the manifold of states represents a new arena for description of physical processes.
%\item [(S3)]
%The ``classical world'' for a given system can be identified with the set of all eigenstates of all observables 
%\end{description}
%The next statement, although not essential in the Letter, is in line with the previous two and with the entire spirit of the approach. It is likely to become significant in the future development.
%\begin{description}
\item [(S3)]
The interactions can be described in terms of the Riemannian metric on the manifold of states. In particular, the states of microscopic particles move along geodesics on the sphere of states furnished with a Riemannian metric. In this sense, the interactions may have a geometric origin.
\end{description}
%More generally, it seems to be possible to formulate the non-relativistic quantum mechanics with measurements as a metric theory on the space of states of the system under consideration. This program has a long way to go and the obtained results are very preliminary. One exciting possibility is  

%Note that in principle such a formulation is capable of putting the measuring device and the measured system on an equal footing. Not only both, the system and the device are represented by points on a Hilbert manifold of states, but

%As already mentioned, the geometric origin of interactions leading to the process of collapse will not be explored in the Letter.

Let us investigate how these statements together with the hypotheses {\bf (H1)},{\bf (H2)} in the previous section help provide an understanding of the measurement process in QM. 
First of all, a particular measuring device can be modeled by a metric field with sources at the eigenstates of the measured observable. That is, the {\em kind} of measurement performed on the system determines a specific field created by the device on the manifold of states of the system. 
Provided the model based on the hypotheses can be developed into a consistent physical theory, the latter result would resolve the so-called {\em preferred basis problem} in QM.
%
%If correct, this would resolve the so-called {\em preferred basis problem} in QM.  
The problem can be formulated as follows: 
\begin{description}
\item [(P1)]
How could the electron's state $\varphi$ ``know'', which basis $\{e_{k}\}$ to use to associate the right probabilities to the coefficients $c_{k}$ in decomposition $\varphi=\sum_{k} c_{k}e_{k}$? 
\end{description}
The constructed model suggests the following answer:
\begin{description}
\item[(R1)]
The coefficients of state of the system in the basis of eigenvectors of the observable describe position of the state relative to the sources of the field created by the device. As already discussed, this position determines the probability for the state to be ``pushed'' by sources to a particular eigenstate point on the projective space of states.
In other words, by creating a surrounding field in the space of states, the device itself defines the ``preferred'' basis.
\end{description}

Next, the process of collapse in the model is an objective process driving the state of the system to an eigenstate of the measured observable. The stochastic nature of the process is due in the model to random fluctuations of sources  associated with measuring ``parts'' of the device. These fluctuations could be directly related to the usual chaotic oscillations of the ``parts'' {\em extended} to the space of states of the system. 

The ``classical world'' in the approach is represented by eigenstates of observables. The set of all eigenstates of an observable ${\widehat A}$ for a quantum system will be called the set of {\em ${\widehat A}$-classical states} (or {\em points}) for the system. 
So the set of ${\widehat A}$-classical states is a subset in the Hilbert space of states (or the corresponding space of physical states) of the system.
%
%For example, the set of all eigenstates of the position observable ${\bf x}$ for a particle in Euclidean space $R^{3}$ is the set $M_{3}$ of all delta functions $\delta_{{\bf a}}({\bf x})\equiv \delta({\bf x}-{\bf a})$ with ${\bf a} \in R^{3}$. So the set of ${\bf x}$-classical points for the particle is the set $M_{3}$ which can be shown to be a three-dimensional submanifold in an appropriate Hilbert space of states of the particle. Such a submanifold can in fact be identified with the Euclidean space $R^{3}$ itself (see Ref. \cite{Kryukov3}).  
%
Let us point out that there is nothing special about the classical states as what is ``classical'' with respect to one observable is ``quantum'' with respect to another one. 
%To put it differently, ``classicality'' is a reference frame dependent notion; it depends on our choice of a reference system on the space of states.  

In the model under investigation, the integral curves of vector fields associated with observables are geodesics in the Killing metric on the sphere of states. More generally, the integral curves of vector fields associated with any reasonable set of physical observables of a quantum system can be shown to be geodesics in an appropriate Riemannian metric on the sphere of states of the system. The non-commutativity of observables is then tied to the curvature of the metric. 
%Under the assumptions of the model the sectional curvature has been found to be $1/\hbar^{2}$. The fact that the curvature is large  would then explain why the measure of non-commutativity of physical observables is so small, that is, why the world is largely ``classical''. For instance, 

Let us investigate in this light the ``mother of quantum mechanics'', i.e., the double-slit experiment. 
There are two main paradoxes associated with the experiment: 
\begin{description}
	\item [(P2)]
	
How could the electron pass trough both slits at once? 

\item [(P3)]

How could a measuring device inserted after the screen with the slits instantaneously change the way in which the electron has passed through the screen?

\end{description}

The Hilbert space of states in the double-slit experiment is infinite-dimensional. It would be helpful to consider at the same time a version of the experiment with a finite dimensional space of states. For this let us return to the motion of electron in a homogeneous magnetic field. Recall that the spin state of the electron evolves in accordance with equation Eq. (\ref{spin_ev}). If the field is directed along the $Y$-axis and the initial state of the electron is
$\left[ 
\begin{array}{c}
1 \\ 
0
\end{array}
\right]
$,
the solution of Eq. (\ref{spin_ev}) is given by 
\begin{equation}
\varphi_{t}=\left[ 
\begin{array}{c}
\cos\frac{\mu B_{0}}{\hbar}t \\ 
\sin\frac{\mu B_{0}}{\hbar}t
\end{array}
\right].
\end{equation}
If $t$ changes, say, between $0$ and $\frac{\pi}{4}\frac{\hbar}{\mu B_{0}}$, then
the process of passing through the field results in a ``splitting'' of the original spin-up eigenstate of the operator ${\widehat \sigma}_{z}$ into a superposition of spin-up and spin-down states. In this respect the experiment is a finite dimensional version of the double-slit experiment in which a localized electron wave packet gets transformed by the screen with the slits into a superposition of two wave packets. 

With this in hand, let us address the above mentioned paradoxes {\bf (P2)} and {\bf (P3)} of the double-slit experiment. Let us call the original double-slit experiment and the experiment with an electron in a homogeneous magnetic field the {\em E1} and {\em E2} experiments respectively.  
The electron in the experiment {\em E2} evolves from the original ``classical'' state  
$\left[ 
\begin{array}{c}
1 \\ 
0
\end{array}
\right]
$,
into superposition 
$\left[ 
\begin{array}{c}
1/{\sqrt 2} \\ 
1/{\sqrt 2}
\end{array}
\right]
$
of two eigenstates of ${\widehat \sigma}_{z}$. During this evolution the $Z$-component of the electron's spin is unknown. The reason for that is clear: for $0 < t \le \frac{\pi}{4}\frac{\hbar}{\mu B_{0}}$, the trajectory of electron's state on $S^{2}=CP^{1}$ does {\em not} pass through the {\em classical states}, i.e., through the eigenstates of ${\widehat \sigma}_{z}$. Classically speaking, one has a paradox here: the electron's intrinsic angular momentum is not defined. Instead, the electron is in a superposition of states of two different angular momenta. In a way, the electron's spin is up and down {\em at the same time}. 

Note however that the state function $\varphi_{t}$ is defined for $t=0$ as well as for $t > 0$. In other words, it describes the {\em classical} and the {\em non-classical} states equally well. Moreover, any (physical) state of the electron is just a point on $S^{2}$. The evolution of the electron's state is just a path $\{\varphi_{t}\}$ with values in $S^{2}$. The classical way of thinking tells us that the electron somehow splits into two parts that evolve along different paths. However, the actual evolution of the electron is most adequately described by a single path $\varphi_{t}$, thereby confirming the statements {\bf (S1)}, {\bf (S2)}.

The situation in experiment {\em E1} is almost identical, although the paradox here is more dramatic as our classical intuition of position is very strong. Again, the intuition tells us that the electron splits into two parts which are passing through different slits. However, the electron's evolution is most adequately described by a path in the space of states. Of course, such a path does not ``split'' and it describes the evolution of electron before  and after the screen with the slits equally well. 
The resolution of the paradox {\bf (P2)} is then as follows: 
\begin{description}
\item [(R2)]

The electron in the experiment {\em E2} is {\em not} in the spin-up and spin-down states at once. Rather, it is in the state that is neither a spin-up, nor a spin-down state. Similarly, the electron in the experiment {\em E1} does {\em not} pass through two slits at once. Rather, it does not pass through the slits {\em at all}! Indeed, for a state to be a spin-up state, for example, it must be at the north pole of the sphere $S^{2}$ of states, which is not the case for the electron's state in the experiment {\em E2} for $t>0$. Similarly, to pass through a slit is to have a state localized at that slit. But the state of the electron after its interaction with the screen in the experiment {\em E1} is {\em not} localized. In other words, the electron (i.e., the electron's state) is located at a point on the space of states that is different from the point at which an electron passing through the slit would have been. To put it figuratively, the electron passes {\em over}  rather than {\em through} the slits.
\end{description}

One can see that the paradox {\bf (P2)} is resolved by considering the motion of electron in the experiments {\em E1}, {\em E2} as happening in the functional space of states rather than on the classical space or on the space of angular momenta. Vaguely speaking, the ``functional'' (i.e., consistent with {\bf (S1)} and {\bf (S2)}) way of thinking makes the paradox disappear.
In light of this, the resolution of the paradox {\bf (P3)} is now immediate: 
\begin{description}
\item[(R3)]
How could a measuring device inserted after the screen change the way the electron has passed through the screen? The answer is: {\em it does not}! If a counter is inserted behind the screen (and sufficiently far from the screen), the process of ``passing through'' the screen is not affected by it. In particular, the counter can be placed {\em after} the electron has already passed ``through'' the screen and this will {\em not} change the history. Indeed, the evolution of electron is described by a path $\varphi_{t}$. If only one slit is open, this path passes through a point in the space of states which is represented by a state function localized at the slit. If, however, both slits are open, the path {\em does not} pass through such a point. This is true independently of any measurement done behind the screen. What the counter does is to change the path $\varphi_{t}$ for larger values of the parameter $t$ so as to produce a state localized at one of the slits in a way discussed in the previous section. As a result, the final state is {\em as if} the electron had passed through only one of the slits. However, no reality should be attached in this case to the event of passing though the slits. Once again, the electron in the experiment {\em E1} does {\em not} pass through the slits. Likewise, the state function $\varphi_{t}$ does {\em not} describe the probability of passing through one of the slits (but only the probability {\em to be found} by one of the slits). Rather, $\varphi_{t}$ itself represents a new ``functional'' reality of the world which is more adequate in QM than the familiar classical reality. 
\end{description}

To summarize, the paradox {\bf (P3)} is resolved by accepting the statements {\bf (S1)} and {\bf (S2)}, i.e., by recognizing the evolution of electron in the space of states as physical (i.e., {\em real}) and by allowing a ``deformation'' of such an evolution in the presence of a measuring device.

Let us finally analyze a measurement performed on a pair of spin-$1/2$ particles. This will give a hint as to how to proceed in more general cases.
As already discussed in section 4, the total quantum mechanical state of the pair is a point in the tensor product of Hilbert spaces of each particle. In particular, the spin state of a pair of electrons is an element of $C^{2}\otimes C^{2}$. A unit normalized state is a point on the unit sphere $S^{7}$ in this four dimensional complex space. Physical spin states of the pair are then points in the complex projective space $CP^{3}$. Note that there may be points in $CP^{3}$ that do not represent a physical state of the pair. In particular, if the total angular momentum of the pair vanishes, the state of the pair can only be of the form $a\varphi_{+}\otimes \psi_{-}+b\varphi_{-}\otimes \psi_{+}$ with $a,b \in C$. Moreover, if the particles are identical, one must have $a=-b$.

The {\em ${\widehat \sigma}_{z}$-classical points} on $CP^{3}$ are the points where both particles have a specific value of the $Z$-component of spin. These points are represented by the products of $\varphi_{\pm}$ and  $\psi_{\pm}$. In the case when the total angular momentum of the pair vanishes, the points are represented by $\varphi_{+}\otimes\psi_{-}$ and $\varphi_{-}\otimes\psi_{+}$. 
The evolution of the pair is now a path with values in $S^{7}$. This path projects down to a path with values on the underlying space $CP^{3}$. 

With these standard ingredients in place, one can analyze now a version of the famous Einstein-Podolsky-Rosen (EPR) paradox in QM:

\begin{description}
\item[(P4)]Given a pair of spin-$1/2$ particles in entangled state $a_{t}\varphi_{+}\otimes \psi_{-}+b_{t}\varphi_{-}\otimes \psi_{+}$, how could it be, that by measuring the $Z$-component of spin of one of them one fixes the $Z$-component of spin of the other one, even if the particles are far apart?
\end{description}

Note that this paradox is similar to the paradox {\bf (P3)}, taking place on the space of states of the pair. Indeed, the resolution of the paradox is almost identical:

\begin{description}
\item[(R4)] Physical reality is described by the path $\varphi_{t}$ with values in the space $S^{7}$ (or $CP^{3}$) of states of the pair. Unless one of the coefficients $a_{t}, b_{t}$ in  $a_{t}\varphi_{+}\otimes \psi_{-}+b_{t}\varphi_{-}\otimes \psi_{+}$ is zero for some $t$, the path $\varphi_{t}$ does not pass through the ${\widehat \sigma}_{z}$-classical points $\varphi_{+}\otimes\psi_{-}$ or $\varphi_{-}\otimes\psi_{+}$. That is, the particles {\em do not} have any $Z$-component of spin. 
To measure the $Z$-component of spin of a particle is to make the path $\varphi_{t}$ pass through one of the ${\widehat \sigma}_{z}$-classical points $\varphi_{+}\otimes\psi_{-}$ or $\varphi_{-}\otimes\psi_{+}$. In this case (and only in this case!) the $Z$-components of spin of both particles are defined and take opposite values. Furthermore, as with a single particle, the interaction with the measuring device is assumed to cause a ``deformation'' of the path $\varphi_{t}$. The resulting path ends up then at one of the classical points via a stochastic process on the space of states.
\end{description}

The full version of the experiment involving a spatial separation of the particles is even more dramatic. How could the second particle at a point $y$ ``know'' about measurement of the $Z$-component of spin performed on the first particle at a distant point $x$? 
\begin{list}{}{}
\item{}
Again, physical reality of the pair is most adequately described  by a path $\varphi_{t}=a_{t}\varphi_{x+}\otimes \psi_{y-}+b_{t}\varphi_{x-}\otimes \psi_{y+}$ in the space $H$ of states of the pair. Here $\varphi_{x+}$ is the spin-up state of the first particle located at $x$ and similarly for the other state functions in $\varphi_{t}$. The classical points in $H$ have the form $\varphi_{x+}\otimes \psi_{y-}$ and $\varphi_{x-}\otimes \psi_{y+}$. If $\varphi_{t}$ does not pass through these points the spin of individual particles is {\em not} defined. Intuitively, we think that if a particle is ``here'' (at a point $x$), then it ought to have all attributes of a ``real'' particle, including spin. But before the measurement is performed, the particle in the experiment is {\em not really here}! Indeed, it is somewhere else on the sphere of states $S^{H}$ in $H$ (or on the corresponding projective space $CP^{H}$). 
So if reality is associated with the state function of the pair, the paradox is resolved.  
\end{list}

But what about this ``spooky action at a distance''? Notice that the new ``functional'' reality does not use it!
Indeed, the equation of geodesics is ``local'' in the space of states $S^{3}$, because it is a differential equation of geodesics on $S^{3}$.
%Indeed, the measuring device may ``curve'' the path $\varphi_{t}$ {\em locally}, i.e. on a neighborhood of the sources created by the device. 
Of course, this locality in the space of states does not preclude a non-locality in the classical space. 
Indeed, what is a point in the space of states may represent a pair of well separated particles in the classical space.
%The distance between the particles in the classical space may be large, but the distance between $\varphi$ and the functions representing individual particles may be small.
Furthermore, what is close in the metric on the space of functions does not have to be close in the metric  on the classical space (see Ref. \cite{Kryukov}). A detailed analysis of this will be, however, a subject for a different paper.
%%%%%%%%%%%%%%%%%%%%%%%INSERT
%Let us summarize the most important observations made so far. With every observable in the model one can associate a vector field on the space of states $S^{3}$. There exists a Riemannian metric on $S^{3}$ such that the integral curves of the vector fields are geodesics in this metric. In particular, solutions of the Schr{\"o}dinger equation are geodesics. The non-commutativity of observables and the uncertainty principle are due to the curvature of the Riemannian metric. The Planck's constant is the radius of the curvature. The space of ``observed'' states is the projective space $CP^{1}=S^{2}$. This space is furnished with the Fubini-Study metric induced by the embedding of $S^{2}$ into the space of states $S^{3}$. Projection of a geodesic on $S^{3}$ onto $S^{2}$ is not in general a geodesic. The classical points on $S^{2}$ are zeros of the projection of the vector field associated with the measured observable. The uncertainty of an observable in a particular state is a measure of separation of the state from the classical points. The probability of transition of a given state to another state depends only on the distance between the states in the Fubini-Study metric. 

\end{document}